\documentclass[lettersize,journal,transmag]{IEEEtran}
\usepackage{enumitem}
\usepackage{hyperref}
\usepackage{algorithmic}
\usepackage{mathtools}

\usepackage{textcomp}
\usepackage{stfloats}
\usepackage{verbatim}
\usepackage{graphicx}

\usepackage{float}

\usepackage{caption}
\usepackage{subcaption}
\usepackage{color, colortbl}
\usepackage{amsmath, amssymb}
\usepackage{multirow, array}
\usepackage{bm,url}
\usepackage{breakurl}
\usepackage[table,xcdraw]{xcolor}
\usepackage{hhline}
\usepackage{rotating}
\usepackage{adjustbox}
\usepackage{tikz}
\usetikzlibrary{patterns}
\usepackage{collcell}
\usepackage{algorithmic}
\usepackage[ruled,vlined]{algorithm2e}

\usepackage{pgfplots}
\usepackage{subfiles} 

\SetKwInput{KwInput}{Input}  
\SetKwInput{KwOutput}{Output} 

\usepackage{amsthm}

\newtheorem{lemma}{Lemma}
\newtheorem{remark}{Remark}

\def\BibTeX{{\rm B\kern-.05em{\sc i\kern-.025em b}\kern-.08em
T\kern-.1667em\lower.7ex\hbox{E}\kern-.125emX}}
\usepackage{balance}

\SetKwInput{KwInput}{Input}  
\SetKwInput{KwOutput}{Output} 

\begin{document}
\title{Teacher-free Latent Self-distillation and Class-separable Representations for Lightweight IoT Attack Detection}

\author{
	\IEEEauthorblockN{
		Phai~Vu~Dinh,
        Diep~N.~Nguyen,
        Dinh~Thai~Hoang,
        Marwan~Krunz,
		Quang~Uy~Nguyen,\\
		Son~Pham~Bao, and
		Eryk~Dutkiewicz
	}

\thanks{

P. V. Dinh, D. N. Nguyen, D. T. Hoang, and E. Dutkiewicz are with the School of 
Electrical and Data Engineering, University of Technology Sydney, Sydney, 
NSW 2007, Australia (e-mail: Phai.D.Vu@student.uts.edu.au, 
\{diep.nguyen, hoang.dinh, eryk.dutkiewicz\}@uts.edu.au).

M. Krunz and P. V. Dinh are with the Department of Electrical and Computer Engineering, 
The University of Arizona, Tucson, AZ, USA (e-mail: \{krunz, dinhphaivu\}@arizona.edu).

N. Q. Uy and P. V. Dinh are with the Institute of Information and Communication Technology, 
Le Quy Don Technical University, Hanoi, Vietnam (email: \{quanguyhn, phaivd\}@lqdtu.edu.vn).

P. B. Son is with the University of Engineering and Technology, 
Vietnam National University, Hanoi, Vietnam (e-mail: sonpb@vnu.edu.vn).

Corresponding author: Phai Vu Dinh

Preliminary results of this work will be presented at the Global Communications Conference in December 2026, Macau, China.
}

}

\markboth{IEEE Internet of Things Journal, No. X, XXXX XXXX}%
{How to Use the IEEEtran \LaTeX \ Templates}

\maketitle

\begin{abstract}
Knowledge distillation (KD) has been widely used to improve lightweight AI models by transferring soft-label knowledge from a large teacher model to a student model. However, existing KD methods are primarily designed for the image domain rather than lightweight IoT devices, and they often struggle to maintain well-separated feature representations for different attack types, especially as the number of classes increases and attack behaviors become more diverse. This paper proposes a novel \textit{teacher-free latent self-distillation framework based on a Twin Autoencoder (TAE)}.
{\color{black}
Instead of relying on an external teacher, TAE generates intrinsic class-wise latent representations through a deterministic transformation of the latent representation, as referred to \emph{class-adaptive target representation (CATR)}, which serves as a target representation similar to soft labels in KD. The decoder then learns to reconstruct the CATR, promoting class separation in the resulting decoder representations. After training TAE, both training and test samples are passed through the decoder to obtain their corresponding representations. These representations are then used as inputs to conventional classifiers, such as Decision Trees (DT), for classification, improving discrimination between benign and attack samples while maintaining a lightweight design suitable for IoT deployment.}~We theoretically derive conditions for perfect class separation and show that lower empirical risk yields better representations, identifying regimes where TAE achieves strictly lower empirical risk than models using fixed class centers. Extensive experiments on 13 cybersecurity datasets, covering IoT botnets, network intrusion detection, malware, cloud DDoS, and synthetic multi-class data, show that TAE achieves up to 96.1\% average accuracy for IoT attack detection and 98.7\% for cloud intrusion detection. With a compact model size (1 MB) and ultra-fast inference (0.26~$\mu$s per sample), TAE offers a practical and scalable solution for real-world cybersecurity and IoT systems.

\end{abstract}

\section{Introduction}

\IEEEPARstart{I}{ntrusion} detection systems (IDS) are essential for safeguarding internet of things (IoT) information systems in the digital age~\cite{IoTJ-Xgb, IoTJ-SD}. Yet, building effective IoT IDS remains challenging due to two key factors. First, learning discriminative representations for IoT intrusion detection becomes increasingly challenging as the diversity and number of attack categories continue to grow. Thousands of new vulnerabilities, malware families, and attack variants emerge each year~\cite{li2021malicious}, expanding the range of behaviors that an IoT IDS must distinguish. As more attack categories are introduced, the boundaries between classes become likely less distinct, since different attacks may generate highly similar network traffic patterns. For example, the IoT dataset~\cite{Meidan_2018} contains more than 1,500 variants of the Mirai botnet, many of which differ only in subtle behavioral details. Moreover, adversaries employ evasion techniques such as encrypted communication and resource-mimicking behaviors, further reducing the distinguishability between attack categories. Consequently, learning feature representations that preserve intra-class consistency (i.e., samples from the same class remain close together) while maintaining strong inter-class separability (i.e., samples from different classes remain well separated) remains a significant challenge for modern IoT IDS. Failure to achieve these properties can reduce classification accuracy and increase false alarm rates. Second, IDS must be deployable not only on servers but also on resource-constrained IoT devices~\cite{10873007}. With limited memory and computational power~\cite{CTVAE}, these devices cannot support high-complexity models, making lightweight yet accurate detection methods critical.

Machine Learning (ML) has been crucial in the success of IDS~\cite{10531267} by learning network traffic patterns to classify new incoming data. However, conventional ML classifiers, such as Decision Trees, Support Vector Machines, and Random Forests, may struggle with the complexity of CDS data \cite{LyVu, CTVAE}, resulting in lower detection accuracy. 
Among ML techniques, representation learning (RL) is a key approach for enabling effective IDS in IoT devices, as it transforms complex input data into abstract features that facilitate downstream attack detection \cite{Goodfellow-book}. However, RL models often struggle with classification tasks with many classes \cite{simonyan2015very}. Very deep networks such as AlexNet, VGG, and ResNet \cite{VGG-Alex-Res}, Transformers (MLP~\cite{Transformer-MLP}, 1DCNN~\cite{Transforer-1DCNN}) can achieve high classification accuracy across numerous classes, but their large sizes make them impractical for IoT devices with limited computational resources \cite{10555405}.

Many RL models rely on the bottleneck layers of an Auto-Encoder (AE) to map the input data into a latent vector at the bottleneck layer \cite{CTVAE}
 \cite{prasad2024augmenting} \cite{LyVu} \cite{spareAE2018} \cite{wu2020vector}. AE-based models may not require millions of parameters and can achieve high attack detection accuracy on IoT devices \cite{CTVAE}. 
Luo et al. \cite{Luo2018CSAEC} proposed a Convolutional Sparse Auto-Encoder (CSAE) that uses the structure of a convolutional AE to separate feature maps for representation learning. The CSAE's encoder serves as a pre-trained model.
The authors in \cite{LyVu} introduced the Multi-Distribution Auto-Encoder (MAE) and Multi-Distribution Variational Auto-Encoder (MVAE), which aim to project benign and malicious data into tightly separated regions by incorporating a penalty in the loss function. CSAE, MAE, and MVAE use regularizer terms to the loss function of AE to constrain and separate the representation of different classes. However, the regularization terms in AE often make training more difficult due to the trade-off with the reconstruction term. 
The authors in \cite{CTVAE} introduced the Constrained Twin Variational AE (CTVAE), which generates data samples of different classes from separable Gaussian distributions in the latent space. Then, CTVAE projects the input data of different classes into these separable Gaussian distributions to obtain data representation.
However, like the variational auto-encoder (VAE) variants, CTVAE generally suffers from the problem of posterior collapse \cite{takida2022preventing}. Posterior collapse occurs when the KL-divergence term in CTVAE's loss reaches zero, leading to a lack of information from the input data used to generate samples of different classes in the latent space. This likely removes the relationship, i.e., Euclidean distance, among the data samples within a class, leading to a distribution that does not follow the input data. This may lower classification accuracy when using the data representation of these variants. Additionally, AE-based models, such as CSAE \cite{Luo2018CSAEC}, MAE, MVAE \cite{LyVu}, and CTVAE \cite{CTVAE}, may struggle with training sets that have many classes. In addition, the conventional fixed codeword used in supervised representation learning models may be ineffective in separating data samples of different classes. These codewords, i.e., one-hot encoding, represent the target vector of a class \cite{evron2023role} \cite{yang2015deep} but have no relation to the characteristics of the training data. 

Knowledge distillation (KD) is widely used to improve lightweight IoT IDS based on representation learning \cite{hinton2015distilling}. KD typically requires training a large teacher model on extensive datasets to transfer soft-label knowledge to IoT clients. Soft labels are the probability distributions over all classes produced by the teacher, which provide richer information than hard class (one-hot encoding) labels by indicating the teacher’s confidence in each class. However, maintaining such large-sized teacher models increases computational cost, introduces multi-stage training complexity, and limits the scalability of IoT student models on resource-constrained devices.

Recently, self-distillation (SD) has gained attention as a technique in machine learning where a model is retrained using its own previous outputs (or intermediate representations) as soft targets to guide and improve its performance. The authors in~\cite{SDSSL} proposed SDSSL, which aims to align the representations of lower layers with the final layer to improve the linear separability of data from different classes. To go beyond this, ATAS in~\cite{ATAS} integrates global-to-local, local-to-local, and global-to-global distillation to refine feature representations across all scales. Similarly, COSMOS in~\cite{COSMOS} constructs global and local input views for SD, while employing a cross-attention module to learn multimodal representations. SDSSL, ATAS, and COSMOS have demonstrated promising performance in image representation tasks, where spatial information is highly relevant. However, IoT data generally lack inherent spatial structure, which may limit the effectiveness of representations learned by these methods for downstream classification tasks. Furthermore, these approaches may not be well-suited to the limited computational capacity of portable IoT devices. 

Particularly, the authors in~\cite{IoTJ-SD} introduced TBCLNN, a lightweight IoT IDS based on self-knowledge distillation. TBCLNN employs the binary Harris Hawk optimization algorithm (bHHO) for dimensionality reduction of traffic features and applies block convolution (TBC) as a lightweight convolutional operation to design a lightweight neural network. Compared with TBCLNN, our work follows a different direction. Instead of applying self-distillation (SD) directly to classification, we focus on SD-based representation learning. Our proposed SD method addresses the increasing complexity of IoT data by learning more separable data representations, particularly when the number of classes becomes extremely large. Unlike conventional deep models that require the entire model during inference, the proposed method uses only a partially lightweight network model to extract compact and discriminative representations. The lightweight representation extractor can then be combined with lightweight conventional classifiers, such as Decision Trees (DT), to achieve high classification accuracy while maintaining low computational complexity and a small model size suitable for IoT devices.

Overall, the above methods often rely on predefined class targets, such as one-hot vectors, fixed codewords, or manually specified latent distributions. These targets are assigned independently of the training data and therefore do not capture the underlying relationships among classes. As the number of attack classes increases, mapping diverse class distributions to predefined targets becomes increasingly difficult, leading to less compact class clusters and reduced separation between different attack classes in the learned feature space.
Consequently, samples from different classes may overlap in the latent space, reducing the effectiveness of downstream classifiers. This limitation motivates the use of self-distillation, where the target representations are learned directly from the data rather than predefined in advance. By generating class-dependent soft labels from its own latent representations, a self-distillation framework can dynamically construct discriminative targets that better reflect the intrinsic structure of the data. Such adaptive soft labels become increasingly valuable as the number of classes grows, enabling improved class separation without requiring a large teacher model or a substantial increase in model complexity.

\begin{table*}[t] 
\centering
\setlength{\tabcolsep}{2pt}  
\renewcommand{\arraystretch}{1.0}  
\small  
\caption{\color{black} Comparison of the proposed TAE model with related works.}
\label{tab:model_discussion-inblack}

\begin{tabular}{|p{3.8cm}|p{4.2cm}|p{4.2cm}|p{4.2cm}|}  
\hline
\multicolumn{1}{|c|}{\multirow{2}{*}{\textbf{\textcolor{black}{Models}}}} & \multicolumn{2}{c|}{\textbf{\textcolor{black}{Data Complexity}}} & \multicolumn{1}{c|}{\multirow{2}{*}{\textbf{\textcolor{black}{IoT Deployment}}}} \\ 
\cline{2-3}
\multicolumn{1}{|c|}{} & \multicolumn{1}{c|}{\textbf{\textcolor{black}{High Dimensions}}} & \multicolumn{1}{c|}{\textbf{\textcolor{black}{Multiple-class problem}}} & \\ 
\hline

\textcolor{black}{SVM, DT, RF \cite{10531267}; XGBoost \cite{xgboost-2023, IoTJ-Xgb}} & \textcolor{black}{Low accuracy \cite{CTVAE}, \cite{LyVu}} & \textcolor{black}{Low accuracy} & \textcolor{black}{Lightweight; suitable for IoT \cite{CTVAE}, \cite{costa2021features}} \\ \hline

\textcolor{black}{MLP \cite{yin2023igrf}, 1D-CNN \cite{1D-CNN}} & \textcolor{black}{Can achieve high accuracy \cite{cherfi2025mlp}} & \textcolor{black}{Performance degrades with multiple-class problem \cite{1D-CNN}} & \textcolor{black}{Suitable for IoT \cite{cherfi2025mlp}} \\ \hline

\textcolor{black}{AlexNet, VGG, ResNet~\cite{VGG-Alex-Res}; MLP Transformer MLP~\cite{Transformer-MLP}, Transformer-1DCNN~\cite{Transforer-1DCNN}, } & {\textcolor{black}{High accuracy \cite{VGG-Alex-Res}}} & {\textcolor{black}{High accuracy \cite{VGG-Alex-Res}}} & \textcolor{black}{Difficult due to size \cite{10555405}} \\ \hline

\textcolor{black}{AE-based models (CSAE \cite{Luo2018CSAEC}; MVAE, MAE \cite{LyVu})} & \textcolor{black}{High accuracy \cite{CTVAE}, \cite{LyVu}} & \textcolor{black}{Struggles with multiple-class problem} & \textcolor{black}{Suitable for IoT \cite{CTVAE}, \cite{LyVu}} \\ \hline

\textcolor{black}{CTVAE \cite{CTVAE}} & \textcolor{black}{High accuracy \cite{CTVAE}} & \textcolor{black}{May struggle due to posterior collapse \cite{takida2022preventing},  \cite{razavi2019preventing}} & \textcolor{black}{Deployable but complex training} \\ \hline

\textcolor{black}{SDSSL~\cite{SDSSL}, ATAS~\cite{ATAS}, COSMOS~\cite{COSMOS}} & \textcolor{black}{May high accuracy \cite{CTVAE}} & \textcolor{black}{Struggles with multiple-class problem} & \textcolor{black}{Difficult due to size} \\ \hline

\textbf{\textcolor{black}{Proposed TAE}} & \multicolumn{2}{l|}{\textbf{\textcolor{black}{High accuracy ($\geq$ 2\% over related models)}}} & \textbf{\textcolor{black}{IoT-friendly, lower training complexity}} \\ \hline
\end{tabular}

\end{table*}

\begin{table}[t]
	\caption{{\color{black} Notation used in the TAE methodology and analysis.}}
	\label{tab:notation}
	\centering
	\setlength{\tabcolsep}{1pt}
	\renewcommand{\arraystretch}{1.15}
	\scriptsize
	\begin{tabular}{|c|p{0.85\linewidth}|}
		\hline
		\textbf{Notation} & \textbf{Definition} \\ \hline
		
		$\mathbf{x}^{(i)}$
		& $i$th input sample \\ \hline
		
		$\mathbf{x}^{(i,c)}$
		& $i$th input sample belonging to class $c$ \\ \hline
		
		$\mathbf{X}$
		& Dataset $\{\mathbf{x}^{(i)}\}_{i=1}^{n}$ \\ \hline
		
		$C$
		& Set of classes, with $|C|$ denoting the number of classes \\ \hline
		
		$n_c$
		& Number of samples belonging to class $c$ \\ \hline
		
		$d$
		& Dimensionality of the input representation \\ \hline
		
		$\mathbf{e}^{(i,c)}$
		& Latent representation of $\mathbf{x}^{(i,c)}$ produced by the TAE encoder \\ \hline
		
		$\boldsymbol{\mu}^{(c)}$
		& Mean latent representation of class $c$ \\ \hline
		
		$\overline{\boldsymbol{\mu}}$
		& Global mean of the class means $\boldsymbol{\mu}^{(c)}$ \\ \hline
		
		
		
		$t_i^{(c)}$
		& Direction indicator for the $i$th component of class $c$, set to $+1$ if $\mu_i^{(c)}\geq\overline{\mu}_i$ and $-1$ otherwise. \\ \hline
		
		$S$
		& Positive hyperparameter controlling the scale of the class-dependent transformation \\ \hline
		
		
		
		$S^{(c)}$
		& Class-dependent transformation scale, defined as
		$S^{(c)}=S k_c$, where $k_c\in\{1,\ldots,m\}$ is the rank of class $c$
		in the descending order of its training sample size. \\ \hline
		
		$\hat{\boldsymbol{\mu}}^{(c)}$
		& Transformed mean of class $c$, defined as $\hat{\boldsymbol{\mu}}^{(c)}=S^{(c)}\mathbf{t}^{(c)}$ \\ \hline
		
		$\vec{\mathbf{v}}^{(c)}$
		& Class-dependent transformation vector connecting $\boldsymbol{\mu}^{(c)}$ and $\hat{\boldsymbol{\mu}}^{(c)}$ \\ \hline
		
		$\mathbf{z}^{(i,c)}$
		& Class-adaptive target representation (CATR) obtained by transforming $\mathbf{e}^{(i,c)}$ using $\vec{\mathbf{v}}^{(c)}$ \\ \hline
		
		$\hat{\mathbf{x}}^{(i)}$
		& Reconstruction of $\mathbf{x}^{(i)}$ generated by the hermaphrodite network from $\mathbf{z}^{(i)}$ \\ \hline
		
		$\hat{\mathbf{z}}^{(i)}$
		& Reconstruction representation obtained by feeding $\mathbf{z}^{(i)}$ through the hermaphrodite and decoder during training \\ \hline
		
		$\hat{\boldsymbol{\zeta}}^{(i)}$
		& Reconstruction representation generated by directly feeding $\mathbf{x}^{(i)}$ to the decoder; this is the representation used during inference \\ \hline
		
		$\boldsymbol{\xi}^{(i,c)}$
		& Final learned representation of class $c$ used in the empirical-risk analysis of TAE \\ \hline
		
		$\bar{\boldsymbol{\xi}}_c$
		& Empirical class center of the TAE representations for class $c$ \\ \hline
		
		$\bar{\mathbf{z}}_c$
		& Empirical class center of the learned representations $\mathbf{z}^{(i,c)}$ \\ \hline
		
		$\Delta$
		& Minimum distance between any two class centers in the representation space \\ \hline
		
		$r_{\max}$
		& Maximum within-class distance from a representation sample to its corresponding class center \\ \hline
		
		$R$
		& Empirical risk computed using the empirical class centers $\bar{\mathbf{z}}_c$ \\ \hline
		
		$R_m$
		& MLP's empirical risk computed using the fixed class centers $\hat{\boldsymbol{\mu}}^{(c)}$ \\ \hline
		
		$R_a$
		& MAE's empirical risk computed using the fixed class centers $\hat{\boldsymbol{\mu}}^{(c)}$ \\ \hline
		
		$R_v$
		& CTVAE's empirical risk computed by the mean squared  distance between  $\hat{\mathbf{z}}^{(i,c)}$ and $\mathbf{z}^{(i,c)}$ \\ \hline
		
		$R_t$
		& TAE's empirical risk computed as the mean squared distance between $\boldsymbol{\xi}^{(i,c)}$ and $\mathbf{z}^{(i,c)}$ \\ \hline

		$R_{\xi}$
		& TAE's empirical risk computed as the mean squared distance between $\boldsymbol{\xi}^{(i,c)}$ and  $\bar{\boldsymbol{\xi}}_c$ \\ \hline
		
		$S_z$
		& Within-class spread of the target representations $\mathbf{z}^{(i,c)}$ \\ \hline

		$S_{\xi}$
		& Within-class spread of the learned representations $\boldsymbol{\xi}^{(i,c)}$ \\ \hline
		
		$E_c$
		& Representation deviation energy between $\boldsymbol{\xi}^{(i,c)}$ and $\mathbf{z}^{(i,c)}$ \\ \hline
		
	\end{tabular}
\end{table}

{\color{black}
In this paper, we propose a \textit{teacher-free latent self-distillation framework based on a Twin Autoencoder (TAE)}. Instead of relying on an external teacher, TAE learns intrinsic class-wise latent representations and uses them as \emph{class-adaptive target representations (CATRs)} to guide the decoder. This design enables discriminative representation learning without the additional overhead of a teacher model.}~Experiments are conducted on a diverse set of cybersecurity datasets, including seven IoT botnet datasets \cite{Meidan_2018}, two network IDS datasets \cite{NSLKDD}, \cite{UNSW}, three malware datasets \cite{IoT-23}, one cloud DDoS dataset \cite{CloudIDS}, and ten artificial datasets as the number of classes in the training set increases. The results demonstrate the superior performance of the proposed method compared with state-of-the-art self-distillation methods (SDSSL~\cite{SDSSL}, ATAS~\cite{ATAS}, COSMOS~\cite{COSMOS}), representation learning (RL) models (CTVAE~\cite{CTVAE}, MAE, MVAE~\cite{LyVu}, CSAEC~\cite{Luo2018CSAEC}), conventional classifiers (XGBoost~\cite{xgboost-2023, IoTJ-Xgb}, SVM, DT, RF, 1D-CNN, and MLP), transformer-based classifiers built upon MLP~\cite{Transformer-MLP} and 1D-CNN~\cite{Transforer-1DCNN}, as well as hard-coding methods, including one-hot encoding, Hadamard codes~\cite{evron2023role, yang2015deep}, and MAE codewords~\cite{LyVu}.
 The comparison of the proposed TAE with related methods is described in Table \ref{tab:model_discussion-inblack}.

 Our major contributions are summarized as follows:
	
\begin{itemize}
    \item We propose a teacher-free latent self-distillation framework based on a TAE. Instead of relying on an explicit teacher model to provide soft labels, the TAE encoder generates {\color{black} CATR} through a deterministic transformation of the latent representations. The decoder then learns to reconstruct the input from these {\color{black} CATRs}, producing discriminative representations for downstream tasks.

    \item We theoretically derive conditions for perfect class separation and show that lower empirical risk yields better representations, identifying regimes where TAE achieves strictly lower empirical risk than models using fixed class centers, i.e., MLP~\cite{carbonell1983ML, cherfi2025mlp} and MAE~\cite{LyVu}. The testing complexity for all three models, i.e., TAE, CTVAE, and MAE, is the same, i.e., $\mathcal{O}\Big(NTd^2\Big)$ which is suitable for lightweight IoT devices. Experimental results indicate that TAE is well-suited for cybersecurity applications and potentially for IoT systems, where the model size is approximately 1 MB, and the average running time for extracting a data sample is around 0.26~$\mu$s.

    \item We show that TAE's {\color {black} CATR} outperform traditional fixed codewords, including one-hot encoding, Hadamard \cite{evron2023role} \cite{yang2015deep}, MAE's codeword \cite{LyVu}, {\color{black}and soft labels in KD using deep MLP and ResNet teachers~\cite{hinton2015distilling},} in enhancing classifiers' accuracy when using TAE's data representations. Using TAE's means as fixed codewords can enhance the classification accuracy of classifiers applied to the data representations of RL models, such as MLP and MAE.

\item We conduct extensive experiments on diverse cybersecurity datasets, including seven IoT botnet datasets~\cite{Meidan_2018}, two network intrusion detection datasets~\cite{NSLKDD, UNSW}, three malware datasets~\cite{IoT-23}, one cloud-based DDoS dataset~\cite{CloudIDS}, and ten synthetic datasets designed to evaluate performance as the number of training classes increases. TAE achieves up to 96.1\% average accuracy for IoT attack detection and 98.7\% for cloud intrusion detection, outperforming related models. Experiments on synthetic datasets further show that TAE effectively maps the original data into a lower-dimensional latent space, where samples from different classes become increasingly separable as the number of classes grows.
\end{itemize}

The remainder of this paper is organized as follows. 
Our proposed architecture and model are described in Section \ref{sec:proposed_methodology}, and theoretical analysis is provided in Section \ref{l:tae_analysis}. Section \ref{sec:experimental_settings} presents the experimental settings. The experimental results and discussion are given in Section \ref{sec:performance_analysis}. The characteristics of TAE are discussed in Section \ref{sec:TAE_analysis}. Finally, Section \ref{sec:conclusion} concludes the paper and outlines future research directions.

\section{Twin Auto-Encoder}
\label{sec:proposed_methodology}



\begin{figure}[t]
    \centering
    \subfloat[TAE architecture.\label{fig:tae_architecture}]{
        \includegraphics[width=0.45\textwidth]{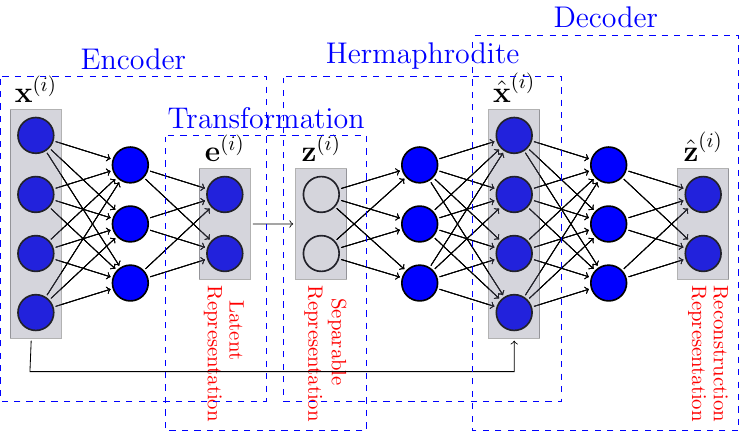}
    }
    \hfill
    \subfloat[{\color{black}Training and inference of TAE.}\label{fig:tae_process}]{
        \includegraphics[width=0.49\textwidth]{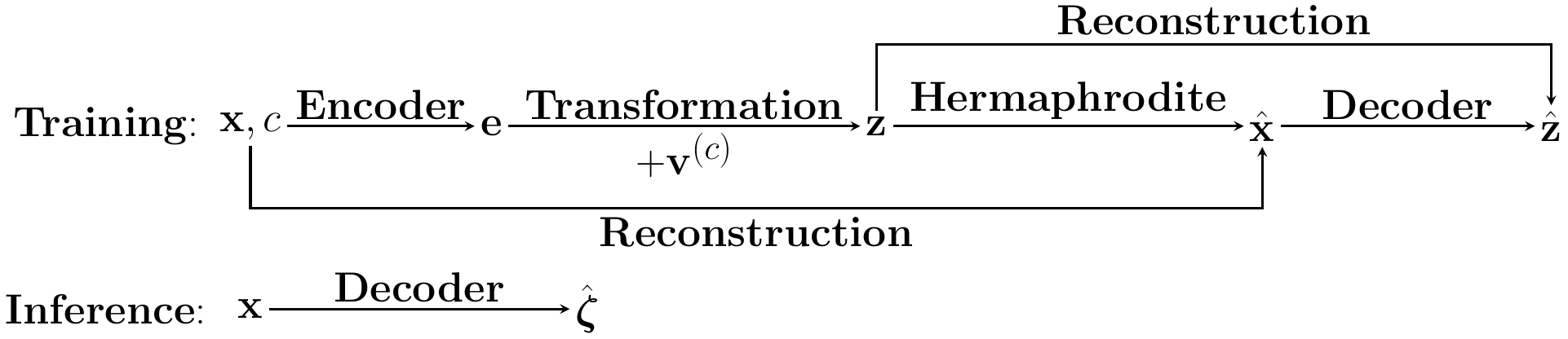}
    }
    \caption{ {\color{black} Overview of the proposed Twin Autoencoder (TAE): 
    (a) TAE architecture for the $i$th input sample, and 
    (b) training and inference processes.}}
    \label{fig:tae_overview}
\end{figure}



\subsection{TAE Architecture}

TAE consists of three networks: An encoder, a hermaphrodite, and a decoder, as illustrated in~Fig.~\ref{fig:tae_overview} (a).  The encoder maps the $i$th input sample $\textbf{x}^{(i)}$ into a latent vector $\textbf{e}^{(i)}$. 
{\color{black}
Unlike conventional fixed codewords, such as one-hot or Hadamard codes, which assign the same predefined vector to every sample within a given class, the proposed \emph{class-adaptive target representation (CATR)} is sample-dependent and is given by:
\begin{equation}
	\mathbf{z}^{(i,c)} \coloneq \mathbf{e}^{(i,c)}+\mathbf{v}^{(c)}
\end{equation}
where $\mathbf{e}^{(i,c)}$ denotes the latent representation of sample $\mathbf{x}^{(i,c)}$ and $\mathbf{v}^{(c)}$ denotes the class-dependent transformation vector for class $c$. Therefore, samples from the same class can have different CATRs while $\mathbf{v}^{(c)}$ varies across classes (see Table~\ref{tab:notation}).

During training, the class label $c$ is used only to construct the CATR through the corresponding transformation vector $\mathbf{v}^{(c)}$. During inference, the class label is unavailable and is not required. Instead, a new input $\mathbf{x}$ is directly fed into the trained decoder to obtain the final reconstruction representation $\hat{\boldsymbol{\zeta}}$ (see ~Fig.~\ref{fig:tae_overview} (b)).
}

The hermaphrodite network connects the encoder and decoder; thus, it plays both decoding and encoding roles. For its decoding role, it reconstructs the input $\textbf{x}^{(i)}$ as $\hat{\textbf{x}}^{(i)}$ using the separable vector $\textbf{z}^{(i)}$. For the encoding role, the hermaphrodite maps the separable vector  $\textbf{z}^{(i)}$ into a new space $\hat{\textbf{x}}^{(i)}$.

The TAE decoder attempts to project the output of the hermaphrodite into the {\color{black} CATR} to obtain $\hat{\textbf{z}}^{(i)}$. This $\hat{\textbf{z}}^{(i)}$ is called the reconstruction representation in the sense that it is reconstructed from the {\color {black} CATR} $\textbf{z}^{(i)}$. The reconstruction representation is used as the final representation of TAE for the next detection models. 


\subsection{Transformation Operator}
\label{l_translation}

\begin{figure}[t]
	\centering
	\includegraphics[width=0.35\textwidth]
	{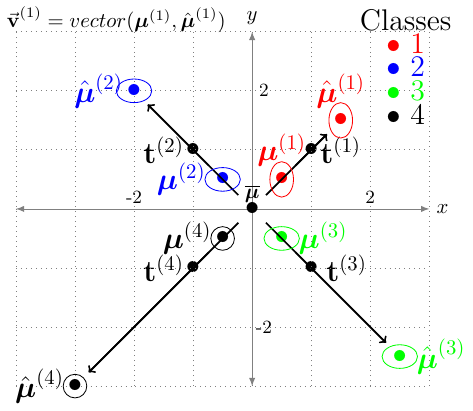}
	\caption{Example of applying the transformation operator in TAE on a 4-class dataset.}
	\label{fig:encoder-moved} 
\end{figure}

\begin{algorithm}[t] 
	\small
	\SetAlgoLined
	\DontPrintSemicolon
	\KwInput{\;

Dataset $\textbf{X}={\{\textbf{x}^{(i)}\}_{i=1}^n}$ where ${\textbf{x}}^{(i)}$ is $i$th samples in class $c$, $C$ is the set of all classes, $n_c$ is the number of samples in class $c \in C$, $|C|$ is the number of classes. ${\textbf{x}}^{(i,c)}$ denotes ${\textbf{x}}^{(i)}$ with label $c$.\\
	}
	\KwOutput{
		$\boldsymbol{\mu}^{(c)}$, $\boldsymbol{\hat{\mu}}^{(c)}$, soft label $\textbf{z}^{(i,c)}$
	}
	\textbf{Step 1}: Apply PCA to reduce the dimension of the input data.\;
	{
		$\textbf{x}_{r}  \gets PCA(X)$ \;
	}
	\textbf{Step 2}: Calculate mean $\boldsymbol{\mu}^{(c)}$ and center mean $\boldsymbol{\overline{\mu}}^{(c)}$ of class $c$. \;
	\ForEach{$c$ in $C$}{
		$\boldsymbol{\mu}^{(c)} \gets \frac{1}{n_c} \sum_{i=1}^{n^c} (\textbf{x}^{(i,c)}_{r})$
	}
	$\boldsymbol{\overline{\mu}} \gets \frac{1}{|C|}\sum_{c=1}^{|C|}(\boldsymbol{\mu}^{(c)})$ \;
	\textbf{Step 3}: Calculate the transformed mean $ \boldsymbol{\hat{\mu}}^{(c)} $ of $\boldsymbol{\mu}^{(c)}$ for each class $c$. \;

	$\boldsymbol{\hat{\mu}}^{(c)} \gets S^{(c)}  \times {\textbf{t}}^{(c)}$
	\;
	\textbf{Step 4}: Transform $\textbf{e}^{(i,c)}$ to a new representation $\textbf{z}^{(i,c)}$. \;
	\ForEach{$\textbf{x}^{(i,c)}$ in $\textbf{X}$}{
		Soft label: $\textbf{z}^{(i,c)} \gets \textbf{e}^{(i,c)} + \vec{\textbf{V}}(\boldsymbol{\mu}^{(c)},\boldsymbol{\hat{\mu}}^{(c)}) $
	}
	\caption{Transforming the latent representation into a separable representation to generate {\color {black} CATR}.}
	\label{algo:separated_multi_cate}
\end{algorithm}


The method to generate {\color {black} CATR} through the transformation operator is exemplified in detail in  Fig.~\ref{fig:encoder-moved} and Algorithm~\ref{algo:separated_multi_cate}. We compute the global mean $\boldsymbol{\overline{\mu}}$ of the entire dataset. Then, for each class $c$, we shift its mean $\boldsymbol{\mu}^{(c)}$ away from $\boldsymbol{\overline{\mu}}$ by moving it further from the origin along the angle bisectors. As explained in algorithm \ref{algo:separated_multi_cate} transforming $\textbf{e}^{(i)}$ to $\textbf{z}^{(i)}$ is done in follow steps: 
\begin{itemize}

\item The first step is to reduce the dimension of the original dataset $X$ to the dimension of the latent vector $\textbf{e}^{(i)}$ using Principal Component Analysis (PCA). 

\item In the second step, the mean $\boldsymbol{\mu}^{(c)}$ of each class $c$ as well as the overall mean $\boldsymbol{\overline{\mu}}$ are calculated. 

\item In the third step, new mean after transforming $\boldsymbol{\hat{\mu}}^{(c)}$ is calculated for of each class $c$ by applying $ \boldsymbol{\hat{\mu}}^c=S^{(c)}  \times \textbf{t}^{(c)}$, where $\textbf{t}^{(c)}$ is the direction to transform from $\boldsymbol{\mu}^{(c)}$ to $\boldsymbol{\hat{\mu}}^{(c)}$. 
{\color{black}
    Let $S^{(c)}=S k_c$, where $k_c\in\{1,\ldots,m\}$ is the rank of class $c$ in the descending order of its training sample size.
	Specifically, classes are ranked in a descending order of their training sample sizes, and $k_c$ is assigned as the corresponding rank. Thus, $k_c=1$ for the class with the largest training sample size and $k_c=m$ for the class with the smallest training sample size. For each feature dimension $i$, $t_i^{(c)}=+1$ when $\mu_i^{(c)}\geq\overline{\mu}_i$, including the equality case, and $t_i^{(c)}=-1$ otherwise. This ordering assigns larger transformation scales to classes with fewer training samples, thereby increasing the separation of their generated target representations and allowing the decoder to better capture the effect of the class-dependent shift during reconstruction.
}


$S^{(c)}=S k_c$, where $k_c\in[2,m+2]$ is determined for each class $c$ according to the descending order of its training sample size, and $S$ is a hyper-parameter that controls the distance between the means of the classes ($S > 0$).

\item The fourth step is to transform the latent vector $\textbf{e}^{(i,c)}$ of class $c$  to a new vector $\textbf{z}^{(i,c)}$ following the direction of vector $\vec{\textbf{v}}^{(c)}$  that connects the old mean $\boldsymbol{\mu}^{(c)}$ to the new mean $\boldsymbol{\hat{\mu}}^{(c)}$, i.e.,  $\vec{\textbf{v}}^{(c)} = \vec{\textbf{V}}(\boldsymbol{\mu}^{(c)},\boldsymbol{\hat{\mu}}^{(c)})$. $\textbf{z}^{(i,c)}$ is referred to as soft label of data sample $\textbf{x}^{(i)}$ with class $c$.
\end{itemize}
{\color{black}
		Unlike conventional knowledge distillation, TAE does not employ a separately trained teacher network to generate target outputs. Instead, TAE uses the encoder representation $\mathbf{e}^{(i,c)}$ together with a deterministic class-dependent transformation, constructed from PCA-based class statistics and mean shifting, to generate the CATR $\mathbf{z}^{(i,c)}=\mathbf{e}^{(i,c)}+\mathbf{v}^{(c)}$. Thus, the encoder and deterministic transformation jointly serve as an implicit target generator. The decoder is subsequently trained to reproduce this encoder-derived target representation from the input, allowing TAE to learn from targets generated internally by its own representation rather than by an external teacher network. We refer to this mechanism as~\emph{teacher-free self-distillation} to emphasize that the target representation is deterministically generated from the model's own representations without requiring a separately trained, high-capacity teacher network or an additional teacher-training stage.
}


\subsection{Loss Function}
TAE is trained in a supervised manner. The loss function of TAE for a dataset $\textbf{X}={\{\textbf{x}^{(i)}\}_{i=1}^n}$ with labels $\textbf{Y}={\{{\textbf{y}}^{(i)}\}_{i=1}^n}$  includes four terms as follows: 
\begin{equation}
\label{eq:loss_TAE}
\begin{aligned}
\ell_{\emph{TAE}}=\frac{1}{n} \sum_{i=1}^{n}\big{[}(\textbf{x}^{(i)}-\hat{\textbf{x}}^{(i)})^2 + (\textbf{z}^{(i)}-\hat{\textbf{z}}^{(i)})^2 \\ + (\textbf{z}^{(i)}- \hat{\boldsymbol{\zeta}}^{(i)})^2 +(\textbf{e}^{(i)}- \boldsymbol{{\mu}}^{(c)})^{2} ]
\end{aligned}
\end{equation}

\noindent where $\hat{\boldsymbol{\zeta}}^{(i)}$ is the reconstruction representation when the input $\textbf{x}^{{(i)}}$ is directly fed into the decoder of TAE. $\boldsymbol{{\mu}}^{(c)}$ the mean of data samples in class $c$ calculated in \textbf{Step 2} of Algorithm \ref{algo:separated_multi_cate}.

The first term in Equation (\ref{eq:loss_TAE}) is the reconstruction error between $\textbf{x}^{(i)}$ and $\hat{\textbf{x}}^{(i)}$. The next  term is reconstruction error between $\textbf{z}^{(i)}$ and $\hat{\textbf{z}}^{(i)}$. The third term is used to train the decoder with the input $\textbf{x}^{(i)}$.
The fourth term aims to shrink the latent vector $\textbf{e}^{(i)}$ into its mean $\boldsymbol{\mu}^{(c)}$ of each class. This term prevents the latent representation $\textbf{z}^{(i)}$ of different classes from being overlapped.

{\color{black}
\section{Conditional Representation Stability Analysis of TAE}
}
\label{l:tae_analysis}

{\color{black}
	
	In this section, we first derive theoretical conditions for perfect representation and characterize the relationship between empirical risk and representation quality.  We then establish the conditional representation stability of the proposed TAE when its empirical error is lower than that of baseline models, such as MLP and MAE, under specific conditions.

}


\subsection{Representation Quality and Conditions for Perfect Separation}

\begin{lemma}[Perfect Representation]
\label{lem_repres_quality}
Define the following quantities:
\begin{equation}
\label{eq:Delta}
    \Delta \coloneqq \min_{a \neq b} \| \bar{\mathbf{z}}_a - \bar{\mathbf{z}}_b \|, \quad a,b \in C,
\end{equation}
\begin{equation}
\label{eq:r_max}
    r_{\max} \coloneqq \max_{c \in C,\, 1 \le i \le n_c} \| \mathbf{z}^{(i,c)} - \bar{\mathbf{z}}_c \|,
\end{equation}
\begin{equation}
\label{eq:9}
    \bar{\mathbf{z}}_c = \frac{1}{n_c} \sum_{i=1}^{n_c} \mathbf{z}^{(i,c)}.
\end{equation}
Then, if $r_{\max} < \frac{\Delta}{2}$, each representation sample $\mathbf{z}^{(i,c)}$ is closer to its own class center $\bar{\mathbf{z}}_c$ than to any other class center $\bar{\mathbf{z}}_{c'}$ for all $c' \ne c$.
\end{lemma}
Lemma~\ref{lem_repres_quality} defines $r_{\max}$ as the largest within-class radius, which measures how far any sample can be from its own class center. $\Delta$ denotes the minimum inter-class distance and measures how far apart the closest pair of class centers are. $\bar{\mathbf{z}}_c$ denotes the class center of the representation data for class $c$ after training.

\begin{proof}
Let $\mathbf{z}$ be a representation belonging to class $a \in C$, and $b\in C$ is another class.  
By the triangle inequality, we have
\begin{equation}
    \| \mathbf{z} - \bar{\mathbf{z}}_b \| 
    \geq \| \bar{\mathbf{z}}_a - \bar{\mathbf{z}}_b \| - \| \mathbf{z} - \bar{\mathbf{z}}_a \|.
\end{equation}
Combining~(\ref{eq:Delta}) and~(\ref{eq:r_max}), it follows that
\begin{equation}
    \| \mathbf{z} - \bar{\mathbf{z}}_b \| \geq \Delta - r_{\max}.
\end{equation}
From~(\ref{eq:r_max}), we also have $\| \mathbf{z} - \bar{\mathbf{z}}_a \| \leq r_{\max}$.  
To ensure that each representation $\mathbf{z}^{(i,a)}$ of class $a$ is closer to its own class center than to any other class center, we require
$\Delta - r_{\max} > r_{\max}$, which is equivalent to $r_{\max} > \frac{\Delta}{2}$.  
Under this condition, the inequality $\| \mathbf{z} - \bar{\mathbf{z}}_b \| > \| \mathbf{z} - \bar{\mathbf{z}}_a \|$ holds for all $b \neq a$.
\end{proof}

\begin{lemma}[Empirical Risk and Representation Quality]
\label{lemma_representation}
Let the empirical risk of a representation model $f$ be defined as
\begin{equation}
\label{eq:emp-risk-center}
R = \frac{1}{N} \sum_{c=1}^{|C|} \sum_{i=1}^{n_c} 
\left\| \mathbf{z}^{(i,c)} - \bar{\mathbf{z}}^{(c)} \right\|^2.
\end{equation}
Then the following statements hold:
1. A smaller empirical risk $R$ indicates that the samples $\mathbf{z}^{(i,c)}$ are closer to their respective class centers, and hence the representation is more compact and discriminative.  
2. If
\begin{equation}
    \label{eq:perfect-condition}
    R < \frac{\Delta^2}{4}.
\end{equation}
then the model $f$ achieves perfect representation.
\end{lemma}
Equation~(\ref{eq:emp-risk-center}) indicates that as $R$ decreases, the representation samples of each class become closer to their corresponding class center, resulting in a better and more compact representation.
Equation~(\ref{eq:perfect-condition}) implies that as $R$ decreases below $\frac{\Delta^2}{4}$, 
the model $f$ achieves a perfect representation, 
in which the representation samples of each class are closer to their own class center 
than to the centers of any other classes.
\begin{proof}
From the definition of~(\ref{eq:r_max}), we have
\begin{equation}
\label{eq:rmax-risk-relation}
\begin{aligned}
\|\mathbf{z}^{(i,c)} - \bar{\mathbf{z}}^{(c)}\|^2 
&\le r_{\max}^2 \\[4pt]
R = \frac{1}{N} \sum_{c,i} 
\|\mathbf{z}^{(i,c)} - \bar{\mathbf{z}}^{(c)}\|^2 
&\le \frac{1}{N} \sum_{c,i} r_{\max}^2 = r_{\max}^2.
\end{aligned}
\end{equation}
From Lemma~\ref{lem_repres_quality}, we have $r_{\max} < \frac{\Delta}{2}$. Therefore, $R < \frac{\Delta^2}{4}$ holds.
\end{proof}

{\color{black}
\subsection{TAE's Conditional Representation Stability}
}

\begin{table}[t]
\caption{{\color{black}Definition of empirical risks of MLP~\cite{carbonell1983ML}, MAE~\cite{LyVu}, CTVAE~\cite{CTVAE}, and TAE.}}
\label{tab:empirical_risks}
\setlength\tabcolsep{3pt}
\centering
\scriptsize
\begin{tabular}{|c|c|}
\hline
\textbf{Model} & \textbf{Empirical Risk Definition} \\ \hline

\centering MLP &
\begin{minipage}[c]{0.8\linewidth}
\centering
\begin{equation}
R_m = \frac{1}{N} \sum_{c=1}^{|C|} \sum_{i=1}^{n_c} 
\left\| \mathbf{m}^{(i,c)} - \hat{\boldsymbol{\mu}}^{(c)} \right\|^2
\label{eq:emp-risk-mlp}
\end{equation}
\end{minipage}
\\ \hline

\centering MAE &
\begin{minipage}[c]{0.8\linewidth}
\centering
\begin{equation}
R_a = \frac{1}{N} \sum_{c=1}^{|C|} \sum_{i=1}^{n_c} 
\left\| \mathbf{a}^{(i,c)} - \hat{\boldsymbol{\mu}}^{(c)} \right\|^2
\label{eq:emp-risk-mae}
\end{equation}
\end{minipage}
\\ \hline

\centering CTVAE &
\begin{minipage}[c]{0.8\linewidth}
\centering
\begin{equation}
R_v = \frac{1}{N} \sum_{c=1}^{|C|} \sum_{i=1}^{n_c} 
\Big(\left\| \hat{\mathbf{z}}^{(i,c)}  - \mathbf{z}^{(i,c)} 
\right\|^2
\Big)
\label{eq:emp-risk-ctvae}
\end{equation}
\end{minipage}
\\ \hline

\centering TAE &
\begin{minipage}[c]{0.8\linewidth}
\centering
\begin{equation}
R_t = \frac{1}{N} \sum_{c=1}^{|C|} \sum_{i=1}^{n_c} 
\Big(\left\| \boldsymbol{\xi}^{(i,c)} - \mathbf{z}^{(i,c)}  \right\|^2
\Big)
\label{eq:emp-risk-tae}
\end{equation}
\end{minipage}
\\ \hline

\end{tabular}
\end{table}

To evaluate the data representations learned by MLP~\cite{carbonell1983ML}, MAE~\cite{LyVu}, CTVAE~\cite{CTVAE}, and TAE after training, we make the following assumptions:(A0) The trainable components of the compared models---the MLP, the encoder of the MAE, and the decoders of the CTVAE and TAE---are architecture-matched, i.e., they have the same number of layers and the same number of neurons in each corresponding layer. They are trained using the same common training settings, including the learning rate, batch size, weight-initialization seed, and optimization algorithm. Model-specific hyperparameters are selected using the same training/validation procedure and fixed before evaluating the final empirical risks.
(A1) All remaining components are fixed during training, including the decoder of the MAE and the encoder and Hermap modules of the CTVAE and TAE. These components are treated as pre-trained modules.
(A2) All models are trained using the same training data.
{\color{black} 
(A3) The flexibility of the class-specific transformation operator $\mathbf{v}^{(c)}$ is controlled by the hyperparameter $S$, which is selected using the training data to ensure sufficient between-class variance for separating samples from different classes.
(A4) For the theoretical comparison, the MLP and MAE are assumed to produce identical latent representations for the same training sample, i.e., $\mathbf{m}^{(i,c)}=\mathbf{a}^{(i,c)}$ for all $i$ and $c$.
(A5) For the comparison with the target representations, the latent representations produced by the MAE are assumed to coincide with the corresponding target representations, i.e., $\mathbf{a}^{(i,c)}=\mathbf{z}^{(i,c)}$ for all $i$ and $c$.}


We define the corresponding empirical risks, which are directly used for extracting data representations, as shown in Table~\ref{tab:empirical_risks}.
Note that both MLP and MAE use the codewords generated by Algorithm~\ref{algo:separated_multi_cate}, i.e., $\boldsymbol{\hat{\mu}}^{(c)}$, whereas CTVAE uses $\boldsymbol{\hat{\mu}}^{(c)}$ to generate data samples from the distribution $\mathcal{N}(\boldsymbol{\hat{\mu}}^{(c)}, I)$. $\mathbf{m}^{(i)}$ and $\mathbf{a}^{(i)}$ represent the data in the latent space of MLP and MAE, respectively, whereas those of CTVAE and TAE are $\hat{\mathbf{z}}^{(i)}$ and $\boldsymbol{\xi}^{(i)}$, respectively.
A lower value of empirical risk indicates that the model has learned better and likely provides a better data representation~\cite{zhang2017understanding}. 

\begin{remark}
\label{tae_remark_1}
Under assumptions (A0)--(A2), and when the class centers $\hat{\boldsymbol{\mu}}^{(c)}$ are computed from the Algorithm~\ref{algo:separated_multi_cate}, it follows from~(\ref{eq:emp-risk-mlp}) and~(\ref{eq:emp-risk-mae}) that $R_m = R_a$.
\end{remark}
\begin{proof}
Since the mapping from the set of per-sample outputs to the class centers is deterministic, and $\mathbf{m}^{(i,c)} = \mathbf{a}^{(i,c)}$ for all $i, c$, it follows that 
$\hat{\boldsymbol{\mu}}^{(c)}$ is identical for both models, i.e., MLP and MAE. Therefore, 
$\|\mathbf{m}^{(i,c)} - \hat{\boldsymbol{\mu}}^{(c)}\|^2 = \|\mathbf{a}^{(i,c)} - \hat{\boldsymbol{\mu}}^{(c)}\|^2$ 
for all $i, c$. Consequently, $R_m = R_a$ holds.
\end{proof}

\begin{lemma}
\label{tae_lemma_1}
Let $R$ in~(\ref{eq:emp-risk-center}) denote the empirical risk computed using the empirical class centers $\bar{\mathbf{z}}_c$, where each $\bar{\mathbf{z}}_c$ is calculated from the final learned representations after training. 
Let $R_m$ in~(\ref{eq:emp-risk-mlp}) and $R_a$ in~(\ref{eq:emp-risk-mae}) denote the empirical risks computed using the fixed target class centers $\hat{\boldsymbol{\mu}}^{(c)}$, which are predetermined before training and remain constant during the learning process. 
The following satisfies:
\begin{equation}
\label{eq:r_lower_R_a}
R \leq R_a \quad \text{and} \quad R \leq R_m.
\end{equation}
\end{lemma}
{\color{black}
	Lemma~\ref{tae_lemma_1} shows that the empirical class center is the optimal fixed representative for minimizing the within-class  representation error. Specifically, for a given set of learned representations, replacing any fixed class representative with the corresponding empirical class center cannot increase the empirical risk. Therefore, $R\leq R_a$ and $R\leq R_m$ follow directly from the optimality of the empirical class centers. This result provides a reference point for analyzing the empirical error of TAE and motivates the subsequent conditional representation-stability analysis in Lemma~\ref{lemma_tae_center}.
}

\begin{proof}
The left-hand side (LHS) of~(\ref{eq:r_lower_R_a}) holds as the following inequality is satisfied:
\begin{equation}
\label{eq:eq20}
    \sum_{i=1}^{n_c} \| \mathbf{z}^{(i,c)} - \bar{\mathbf{z}}_c \|^2
    \;\le\;
    \sum_{i=1}^{n_c} \| \mathbf{a}^{(i,c)} - \hat{\boldsymbol{\mu}}^{{c}} \|^2.
\end{equation}
Under assumptions (A0)--(A2), we have $\mathbf{a}^{(i,c)} = \mathbf{z}^{(i,c)}$. Let's calculate the right-hand side (RHS) of~\ref{eq:eq20}:
\begin{equation}
\label{eq:21}
\begin{aligned}
    \sum_{i=1}^{n_c} \| \mathbf{a}^{(i,c)} - \hat{\boldsymbol{\mu}}^{(c)} \|^2 
    &= \sum_{i=1}^{n_c} \| (\mathbf{a}^{(i,c)} - \bar{\mathbf{z}}_c ) +  (\bar{\mathbf{z}}_c - \hat{\boldsymbol{\mu}}^{(c)} )\|^2 \\
    &= \sum_{i=1}^{n_c} \| \mathbf{a}^{(i,c)} - \bar{\mathbf{z}}_c \|^2 
    + n_c \| \bar{\mathbf{z}}_c - \hat{\boldsymbol{\mu}}^{(c)} \|^2  \\ 
    &\quad + 2(\bar{\mathbf{z}}_c - \hat{\boldsymbol{\mu}}^{(c)})^{T}
    \sum_{i=1}^{n_c} (\mathbf{a}^{(i,c)} - \bar{\mathbf{z}}_c).
\end{aligned}
\end{equation}
From~(\ref{eq:21}) and (\ref{eq:9}), we have 
\begin{equation}
\begin{aligned}
\label{eq:sum_zi_0}
    \sum_{i=1}^{n_c} (\mathbf{a}^{(i,c)} - \bar{\mathbf{z}}_c) = \sum_{i=1}^{n_c} (\mathbf{z}^{(i,c)} - \bar{\mathbf{z}}_c) = \sum_{i=1}^{n_c} \mathbf{z}^{(i,c)} - n_c \bar{\mathbf{z}}_c =0.
\end{aligned}
\end{equation}
Therefore, from~(\ref{eq:21}) we have
\begin{equation}
\label{eq:23}
\begin{aligned}
    \sum_{i=1}^{n_c} \| \mathbf{a}^{(i,c)} - \hat{\boldsymbol{\mu}}^{(c)} \|^2 
    &= \sum_{i=1}^{n_c} \| \mathbf{a}^{(i,c)} - \bar{\mathbf{z}}_c \|^2 
    + n_c \| \bar{\mathbf{z}}_c - \hat{\boldsymbol{\mu}}^{(c)} \|^2 \\
    &\geq \sum_{i=1}^{n_c} \| \mathbf{z}^{(i,c)} - \bar{\mathbf{z}}_c \|^2 \quad \forall n_c \geq 0.
\end{aligned}
\end{equation}
From remark~\ref{tae_remark_1}, RHS of~(\ref{eq:r_lower_R_a}) also holds.

\end{proof}

In fact, the exact values of $\bar{\mathbf{z}}_c$ are unknown before the training process. However, if we assume that $\bar{\mathbf{z}}_c$ are determined before training, then data samples that are far from their corresponding centers contribute more to the empirical risk, whereas those close to their centers contribute less. Therefore, TAE aims to project data onto {\color{black} CATRs} that are adaptively calculated for each data sample rather than relying on fixed class centers. This approach may lead to a lower empirical risk.

{\color{black}
	
	\begin{lemma} [Conditional Representation Stability of TAE]
		\label{lemma_tae_center}
		Let the empirical risk of TAE be defined with the target representations 
		$\mathbf{z}^{(i,c)}$ fixed as the class centers:
		\begin{equation}
			\begin{aligned}
				R_\xi 
				&= \frac{1}{N} \sum_{c=1}^{|C|} \sum_{i=1}^{n_c} 
				\left\| \boldsymbol{\xi}^{(i,c)} - \bar{\boldsymbol{\xi}}_c \right\|^2,~\text{where}~  
				\bar{\boldsymbol{\xi}}_c = \frac{1}{n_c} \sum_{i=1}^{n_c} \boldsymbol{\xi}^{(i,c)}.
			\end{aligned}
			\label{eq:emp-risk-tae-fix}
		\end{equation}
		For a fixed class $c$, define
		\begin{equation}
			\label{eq:define_delta_z}
			\mathbf{z}_i \coloneqq \mathbf{z}^{(i,c)}, 
			\boldsymbol{\xi}_i \coloneqq \boldsymbol{\xi}^{(i,c)}, 
			\boldsymbol{\delta}_i \coloneqq \boldsymbol{\xi}_i-\mathbf{z}_i, 		\bar{\boldsymbol{\delta}}_c
			\coloneqq
			\bar{\boldsymbol{\xi}}_c-\bar{\mathbf{z}}_c.
		\end{equation}
		Let the within-class target-representation spread, within-class learned-representation spread, and representation deviation energy for class $c$ be defined
		\begin{equation}
			\label{eq:Sz_Sxi}
			\begin{aligned}
				S_z
				&\coloneqq
				\sum_{i=1}^{n_c}
				\left\|
				\mathbf{z}_i-\bar{\mathbf{z}}_c
				\right\|^2,~
				S_{\xi}
				&\coloneqq
				\sum_{i=1}^{n_c}
				\left\|
				\boldsymbol{\xi}_i-\bar{\boldsymbol{\xi}}_c
				\right\|^2,~
				E_c
				&\coloneqq
				\sum_{i=1}^{n_c}
				\left\|
				\boldsymbol{\delta}_i
				\right\|^2.
			\end{aligned}
		\end{equation}
If the deviation energy between the learned and target representations satisfies
\begin{equation}
	\label{eq:Rt_R_xi}
	E_c
	\leq
	(\sqrt{2}-1)^2 S_z,
\end{equation}
then the following representation-stability relation holds:
\begin{equation}
	\label{eq:Rt_lower_Xi}
	R_t \leq R_\xi.
\end{equation}

\end{lemma} 
Lemma~\ref{lemma_tae_center} establishes a sufficient condition for the conditional representation stability of TAE. Specifically, when the deviation energy of the learned representations from their target representations is sufficiently small relative to the intra-class spread of the target representations, the representation deviation is bounded by the within-class spread of the learned representations. Under this condition, the resulting inequality $R_t\leq R_{\xi}$ follows as a consequence of the stability condition.
}

\begin{proof}
    From~(\ref{eq:define_delta_z}), let's calculate as:
    \begin{equation}
 \boldsymbol{\xi}_i - \bar{\boldsymbol{\xi}}_c = (\boldsymbol{\delta}_i + \mathbf{z}_i) -(\bar{\boldsymbol{\delta}}_c + \bar{\mathbf{z}}_c) = (\boldsymbol{\delta}_i-\bar{\boldsymbol{\delta}}_c) + (\mathbf{z}_i-\bar{\mathbf{z}}_c).
     \end{equation}
From~(\ref{eq:Sz_Sxi}), we have:
\begin{equation}
\label{eq:S_xi_2}
    S_{\xi} = \sum_{i=1}^{n_c} \| \boldsymbol{\delta}_i - \bar{\boldsymbol{\delta}}_c \|^2 + \sum_{i=1}^{n_c} \| \mathbf{z}_i - \bar{\mathbf{z}}_c \|^2 + 2 \sum_{i=1}^{n_c} \langle \boldsymbol{\delta}_i - \bar{\boldsymbol{\delta}}_c, \mathbf{z}_i - \bar{\mathbf{z}}_c \rangle.
\end{equation}
Process the first term of~(\ref{eq:S_xi_2}), we have:
\begin{equation}
\label{eq:first_tem}
    \begin{aligned}
    A \coloneqq \sum_{i=1}^{n_c} \|\boldsymbol{\delta}_i - \bar{\boldsymbol{\delta}}_c\|^2
&= \sum_{i=1}^{n_c} \|\boldsymbol{\delta}_i\|^2 
   - 2 \sum_{i=1}^{n_c} \boldsymbol{\delta}_i^\top \bar{\boldsymbol{\delta}}_c
   + n_c \|\bar{\boldsymbol{\delta}}_c\|^2.
    \end{aligned}
\end{equation}
Since $\displaystyle \bar{\boldsymbol{\delta}}_c = \frac{1}{n_c}\sum_{i=1}^{n_c}\boldsymbol{\delta}_i$, we have
\begin{equation}
    \sum_{i=1}^{n_c} \boldsymbol{\delta}_i^\top \bar{\boldsymbol{\delta}}_c
= \bar{\boldsymbol{\delta}}^\top_c \sum_{i=1}^{n_c} \boldsymbol{\delta}_i
= \bar{\boldsymbol{\delta}}^\top_c (n_c \bar{\boldsymbol{\delta}}_c)
= n_c \|\bar{\boldsymbol{\delta}}_c\|^2.
\end{equation}
Combine~(\ref{eq:Sz_Sxi}) and~(\ref{eq:first_tem}), we have
\begin{equation}
\label{eq:A_3}
    A = \sum_{i=1}^{n_c} \|\boldsymbol{\delta}_i\|^2 - n_c \|\bar{\boldsymbol{\delta}}_c\|^2 = S_t - n_c \|\bar{\boldsymbol{\delta}}_c\|^2.
\end{equation}
Process the third term of~(\ref{eq:S_xi_2}), we have:
\begin{equation}
\label{eq:cross_tem}
    \begin{aligned}
B &\coloneqq \sum_{i=1}^{n_c} \langle \boldsymbol{\delta}_i - \bar{\boldsymbol{\delta}}_c, \mathbf{z}_i - \bar{\mathbf{z}}_c \rangle \\
 &= \sum_{i=1}^{n_c} \langle \boldsymbol{\delta}_i, \mathbf{z}_i - \bar{\mathbf{z}}_c \rangle 
 - \sum_{i=1}^{n_c} \langle \bar{\boldsymbol{\delta}}_c, \mathbf{z}_i - \bar{\mathbf{z}}_c \rangle.
    \end{aligned}
\end{equation}
From~(\ref{eq:sum_zi_0}), we have \( \sum_{i=1}^{n_c} (\mathbf{z}_i - \bar{\mathbf{z}}_c) = 0 \). Therefore,
\begin{equation}
\label{eq:zi_0}
 \sum_{i=1}^{n_c} \langle \bar{\boldsymbol{\delta}}_c, \mathbf{z}_i - \bar{\mathbf{z}}_c \rangle = \langle \bar{\boldsymbol{\delta}}_c, \sum_{i=1}^{n_c} (\mathbf{z}_i - \bar{\mathbf{z}}_c) \rangle = \langle \bar{\boldsymbol{\delta}}_c, 0 \rangle = 0.
\end{equation}
Constituting~(\ref{eq:zi_0}) to (\ref{eq:cross_tem}) to obtain:
\begin{equation}
    \label{eq:B_2}
    B =  \sum_{i=1}^{n_c} \langle \boldsymbol{\delta}_i, \mathbf{z}_i - \bar{\mathbf{z}}_c \rangle 
\end{equation}
Combining~(\ref{eq:A_3}), (\ref{eq:B_2}), and~(\ref{eq:Sz_Sxi}), we can rewrite~(\ref{eq:S_xi_2}) as follows:
\begin{align}
    S_{\xi} &= S_t - n_c \|\bar{\boldsymbol{\delta}}_c\|^2 + S_z + 2B, \label{eq:Sxi_expand} \\
    \sigma \coloneqq S_{\xi} - S_t &= S_z + 2B - n_c \|\bar{\boldsymbol{\delta}}_c\|^2. \label{eq:Sxi_minus_St}
\end{align}
Applying the Cauchy–Schwarz inequality to $|B|$, we have:
\begin{align}
    |B| = \left| \sum_{i=1}^{n_c} \langle \boldsymbol{\delta}_i, \mathbf{z}_i - \bar{\mathbf{z}}_c \rangle \right| 
    &\leq \sqrt{\sum_{i=1}^{n_c} \|\boldsymbol{\delta}_i\|^2}
\sqrt{\sum_{i=1}^{n_c} \|\mathbf{z}_i - \bar{\mathbf{z}}_c \|^2} \label{eq:B_3} \\
    |B| &\leq \sqrt{E_c}\, \sqrt{S_z} \label{eq:B_4} \\
    hence \quad B &\geq -\sqrt{E_c S_z} \label{eq:B_5}.
\end{align}
Applying the Cauchy–Schwarz inequality to
$n_c \|\bar{\boldsymbol{\delta}}_c\|^2 
= \frac{1}{n_c} \left\| \sum_{i=1}^{n_c} \boldsymbol{\delta}_i \right\|^2$,
we have
\begin{align}
    n_c \|\bar{\boldsymbol{\delta}}_c\|^2
    &\leq \frac{1}{n_c} \left( n_c \sum_{i=1}^{n_c} \|\boldsymbol{\delta}_i\|^2 \right)
    = \sum_{i=1}^{n_c} \|\boldsymbol{\delta}_i\|^2
    = E_c, \label{eq:delta_mean_bound_corrected_1} \\
    - n_c \|\bar{\boldsymbol{\delta}}_c\|^2 
    &\geq - E_c. \label{eq:delta_mean_bound_corrected_2}
\end{align}
Combining~(\ref{eq:delta_mean_bound_corrected_2}) and~(\ref{eq:B_5}) with~(\ref{eq:Sxi_minus_St}), for $S_z \geq 0$ we obtain:
\begin{align}
    \sigma 
    &\geq S_z - 2\sqrt{E_c S_z} - E_c 
    = S_z \left( 1 - 2\sqrt{\frac{E_c}{S_z}} - \left(\sqrt{\frac{E_c}{S_z}}\right)^2 \right).
\end{align}
Let $ t\coloneqq \sqrt{\frac{E_c}{S_z}} \geq 0$. Equation~(\ref{eq:Rt_lower_Xi}) holds if the following holds:
\begin{align}
f(t) \coloneqq (1-2t-t^2) \geq 0
\end{align}
Solve the corresponding quadratic equation $t^2+2t-1 =0$ to obtain: $t = \frac{-2 \pm \sqrt{(2)^2 - 4(1)(-1)}}{2} 
= \frac{-2 \pm \sqrt{8}}{2} 
= -1 \pm \sqrt{2}.$
For $f(t) \geq 0$, we have $t \in [-1 - \sqrt{2}, -1 + \sqrt{2}]$.
Since $t \geq 0$, it follows that $t \in [0, -1 + \sqrt{2}]$.
Therefore, $E_c \leq (-1 + \sqrt{2})^2 S_z$.

\end{proof}

\subsection{TAE's Training/Inference Complexity}

Assume that the numbers of neurons in the hidden layers of the encoder for TAE, CTVAE \cite{CTVAE}, and MAE \cite{LyVu} are equal, denoted as $\{ \alpha_1 d, \alpha_2 d, \ldots, \alpha_T d\}$, where $d$ is the dimensionality of the input and $T$ is the number of hidden layers in the encoder. Note that $\alpha_i d$ must be an integer for this setting, and $0 < \alpha_i <1$, for $i=\{1,\ldots, T\}$. The training complexity of the TAE encoder is $\mathcal{O}\Big(2NTd^2\Big)$.
The complexity of Principal Component Analysis (PCA) in step 1 of Algorithm \ref{algo:separated_multi_cate} is $\mathcal{O}\Big(Nd^2 + d^3\Big)$ \cite{jolliffe2002principal}.
Due to the symmetry of the TAE, the complexity of training the encoder is the same as that of the hermaphrodite and the decoder. The training complexity of TAE is $\mathcal{O}\Big( 6NTd^2 + Nd^2 + d^3 + 6Nd + 2Cd\Big)$. Therefore, the overall training complexity of TAE is $\mathcal{O}\Big(6NTd^2 + Nd^2 + d^3 + 6Nd + 2Cd\Big)$. Compared with CTVAE~\cite{CTVAE} and MAE~\cite{LyVu}, both TAE and CTVAE use only the decoder to extract representations, similar to how the encoder is used in MAE. Therefore, the testing complexity of all three models is the same, $\mathcal{O}\big(NTd^2\big)$ (see Table~\ref{tab:complexity-TAE-MAE-CTVAE}). This indicates that, during inference, TAE relies solely on the decoder, making it suitable for lightweight IoT devices.

\begin{table}[t]
\caption{{\color{black}Comparison of the computational complexity of TAE, MAE~\cite{LyVu}, and CTVAE~\cite{CTVAE}.}}
\label{tab:complexity-TAE-MAE-CTVAE}
\setlength\tabcolsep{1pt}
\centering
\scriptsize
\begin{tabular}{|c|c|c|}
\hline
\centering
\textbf{} & \centering{{MAE}} 
& \begin{minipage}[c]{0.75\linewidth}
    \centering
    \begin{equation}
      \mathcal{O}\Big( 4NTd^2 + 2Nd \Big)
    \end{equation}
  \end{minipage} \\ \cline{2-3}
\centering
\textbf{} & \centering{{CTVAE}} 
& \begin{minipage}[c]{0.75\linewidth}
    \centering
    \begin{equation}
      \mathcal{O}\Big( 2NTd^2 + 4LNTd^2 + Nd^2 + d^3 + 4Nd + 3Cd \Big)
    \end{equation}
  \end{minipage} \\ \cline{2-3}
\multirow{-3}{*}{\centering\textbf{Training}} 
& \centering{{TAE}} 
& \begin{minipage}[c]{0.75\linewidth}
    \centering
    \begin{equation}
      \mathcal{O}\Big( 6NTd^2 + Nd^2 + d^3 + 6Nd + 2Cd \Big)
    \end{equation}
  \end{minipage} \\ \hline
\centering\textbf{Testing} 
& \centering{{\begin{tabular}[c]{@{}c@{}}TAE,\\ MAE,\\ CTVAE\end{tabular}}} 
& \begin{minipage}[c]{0.75\linewidth}
    \centering
    \begin{equation}
      \mathcal{O}\Big( NTd^2 \Big)
    \end{equation}
  \end{minipage} \\ \hline
\end{tabular}
\end{table}


\section{Experimental settings}
\label{sec:experimental_settings}

\subsection{Datasets}

We use a wide range of datasets in cybersecurity to evaluate the effectiveness of TAE. The tested datasets include the IoT botnet {\color{black}attack datasets}, the network IDS datasets, the cloud DDoS attack dataset, and the malware datasets. The number of training, validation, and testing samples of each dataset are presented in Table~\ref{tab:tab_datasets_classified}. The number of classes, the number of features, and the shorthand (ID)  of each dataset are also shown in this Table. 





The IoT botnet datasets~\cite{Meidan_2018} include seven different datasets. The datasets consist of two  types of attacks, i.e., Gafgyt and Mirai attacks.  We perform two scenarios on these multiclass datasets. The first scenario includes the normal traffic and five types of the Gafgyt attacks that form the six classes classification problem, whilst the second scenario consists of the normal traffic, five types of the Mirai attacks, and five types of the Gafgyt attacks forming the 11 classes problem. The dataset name is shorthanded by the name of the dataset and the type of attacks. For instance, EcoG means the Ecobee dataset and the Gafgyt attack, while EcoMG is the Ecobee dataset with both Mirai and Gafgyt attacks. 




The IDS datasets consist of the NSLKDD (NSL)\cite{NSLKDD} and the UNSW-NB15 (UNSW) \cite{UNSW} dataset. The data in the NSL includes the benign traffic and four types of attacks, i.e., DoS, R2L, Probe, and U2L. Among the four types of attacks, R2L and U2L are two sophisticated and rare attacks that are very challenging to be detected by machine learning algorithms~\cite{yu2020intrusion}. The UNSW dataset has benign traffic and nine types of attacks. Both NSL and UNSW datasets are imbalanced datasets. This is because the number of samples of R2L and U2R attacks in NSL and Worms, Backdoor, and Shellcode attacks in UNSW are significantly lower than those of others. 

\begin{table}[t]
	\caption{Datasets information.}
	\label{tab:tab_datasets_classified}
	\centering
\setlength\tabcolsep{1.5pt}
	\begin{tabular}{|c|c|c|c|c|c|c|}
		\hline
		\multirow{2}{*}{\textbf{Datasets}}& \multirow{2}{*}{\textbf{ID}} & \multirow{2}{*}{\textbf{\begin{tabular}[c]{@{}c@{}}No. \\ Train\end{tabular}}} & \multirow{2}{*}{\textbf{\begin{tabular}[c]{@{}c@{}}No. \\ Valid\end{tabular}}} & \multirow{2}{*}{\textbf{\begin{tabular}[c]{@{}c@{}}No. \\ Test\end{tabular}}} & \multirow{2}{*}{\textbf{\begin{tabular}[c]{@{}c@{}}No. \\ Classes\end{tabular}}} & \multirow{2}{*}{\textbf{\begin{tabular}[c]{@{}c@{}}No. \\ Features\end{tabular}}} \\
		& & & &&&\\ \hline
		DanG \cite{Meidan_2018}& IoT1& 179434 & 76901  & 109863& 6 & 115\\ \hline
		EcoG \cite{Meidan_2018}& IoT2& 158631 & 67986  & 97126 & 6 & 115\\ \hline
		PhiG \cite{Meidan_2018}& IoT3& 239099 & 102472 & 146392& 6 & 115\\ \hline
		737G \cite{Meidan_2018}& IoT4& 192200 & 82372  & 117678& 6 & 115\\ \hline
		838G \cite{Meidan_2018}& IoT5& 199698 & 85586  & 122270& 6 & 115\\ \hline
		EcoMG \cite{Meidan_2018}& IoT6& 409574 & 175533 & 250769& 11& 115\\ \hline
		838MG \cite{Meidan_2018}& IoT7& 410071 & 175746 & 251074& 11& 115\\ \hline
		NSLKDD \cite{NSLKDD}& NSL & 88181  & 37792  & 22544 & 5 & 41 \\ \hline
		UNSW-NB15 \cite{UNSW}& UNSW& 57632  & 24700  & 175341& 10& 42 \\ \hline
		IoT Malware 1 \cite{IoT-23} & Mal1  & 57544  & 14387  & 8888  & 2 & 32 \\ \hline
		IoT Malware 2 \cite{IoT-23} & Mal2  & 6498& 1625& 2603  & 2 & 32 \\ \hline
		IoT Malware 3 \cite{IoT-23} & Mal3  & 10183  & 2546& 53624 & 2 & 32 \\ \hline
		Cloud IDS \cite{CloudIDS}& Cloud  & 1239381& 517915 & 887113& 7 & 24 \\ \hline
	\end{tabular}

\end{table} 

	\begin{table}[t]
		\caption{Grid search settings for hyper-parameters.}
		\label{tab:grid_search_params}
		\setlength\tabcolsep{2.5pt}
		\centering
		\scriptsize
		\footnotesize
		\begin{adjustbox}{max width=\textwidth}
			\begin{tabular}{|c|c|}
				\hline
      KNN& $neighbor$=\{1, 2, 5, 10, 15, 35, 50\} \\ \hline
				SVM& \begin{tabular}[c]{@{}c@{}}LinearSVC: $C$=\{0.1, 0.2, 0.5, 1.0, 5.0, 10.0\} \\ SVC: $decision\_function\_shape={ovo}$\end{tabular}\\ \hline
				DT& \begin{tabular}[c]{@{}c@{}} $max\_depth$=\{5, 10, 20, 50, 100\}\end{tabular}   \\ \hline
				RF& \begin{tabular}[c]{@{}c@{}}  $n\_estimators$=\{5, 10, 20, 50, 100, 150\}\end{tabular}  \\ \hline
				Xgb~\cite{xgboost-2023, IoTJ-Xgb} & \begin{tabular}[c]{@{}c@{}}$min\_child\_weight$=\{5, 10\},  \\ $gamma$=\{0.5, 1.5\}, $subsample$=\{1.0, 0.9\} \end{tabular} \\ \hline
  CSAEC~\cite{Luo2018CSAEC} & 1D-CNN with two Conv1D, $kernel\_size=3$
  \\ \hline
  {\color[HTML]{000000} MAE~\cite{LyVu}} & {\color[HTML]{000000} \begin{tabular}[c]{@{}l@{}}$D_z=\{d_z, 1.5d_z, 2d_z, 3d_z\}$\end{tabular}} \\ \hline
  {\color[HTML]{000000} MVAE~\cite{LyVu}} & {\color[HTML]{000000} \begin{tabular}[c]{@{}l@{}}$D_z=\{d_z, 1.5d_z, 2d_z, 3d_z\}$\end{tabular}} \\ \hline
{\color[HTML]{000000} CTVAE~\cite{CTVAE}}  & {\color[HTML]{000000} $D_z=\{d_z, 1.5d_z, 2d_z, 3d_z\}; S= \{1, 5, 10\}$ } \\ \hline
SDSSL~\cite{SDSSL}  & $D_z=\{d_z, 2d_z\}$; $std= \{0.1, 0.05\}$; $shift= \{1, 5\}$   \\ \hline
ATAS~\cite{ATAS}  & $D_z=\{d_z, 2d_z\}; \beta_1= \{0.1, 0.5, 1\}$; $\beta_2= \{0.1, 0.5, 1\}$  \\ \hline

COSMOS\cite{COSMOS}  & $D_z=\{2d_z\}; \beta_1= \{0.25, 0.5, 1\}$; $\beta_2= \{0.25, 0.5, 1\}$  \\ \hline

Transformer & \\ 
(MLP~\cite{Transformer-MLP}, & \\ 
1DCNN~\cite{Transforer-1DCNN}) & 
\multirow{-3}{*}{$n_{\text{head}}=\{2,3,4\}$} \\ 
\hline

TAE & $D_z=\{d_z, 1.5d_z, 2d_z, 3d_z\}$; $S=\{1, 5, 10, 20\}$
  \\ \hline
  
			\end{tabular}
		\end{adjustbox}
		
	\end{table}


The malware datasets include 20 malware families on IoT systems \cite{IoT-23}. There are three binary malware datasets, e.g., Mal1, Mal2, and Mal3. Each data sample has 32 features. The training sets include benign samples and typical malware families, e.g., DDoS and Part-of-Horizontal-Port-Scan while the testing sets consist of benign samples and more sophisticated malware classes, e.g., C\&C-attack, Okiru, C\&C-heartbeat, C\&C-Mirai, and C\&C-PortScan. In other words, some malware in the testing sets is not available in the training set. Thus, it is difficult for ML models to detect this unknown malware in the testing set. 

The cloud IDS datasets are DDoS attacks on the cloud environment. They are published by Kumar et al.~\cite{CloudIDS} by stimulating the cloud environment and using an open-source platform to launch 15 normal VMs, 1 target VM, and 3 malicious VMs. The extracted datasets contain 24 time-based traffic flow features. The cloud IDS datasets are also divided into training and testing sets with a rate of 70:30. 

{\color{black}
To evaluate the performance of TAE in separating classes as the number of classes in the training set increases, we generate artificial datasets. We use libraries to create datasets with three parameters: the number of classes $n\_classes$, the dimensionality of a data sample $n\_feature$, and the number of data samples per class $n\_points\_per\_class$~\cite{scikit_make_blobs}. We set the $seed$ to 2021 for reproducibility. The dataset is then split into two parts with a 70:30 ratio for training and testing sets. Specifically, ten datasets are created with $n\_classes =\{50, 100, 150, 200, 250, 300, 350, 400, 450, 500 \}$, $n\_feature= 100$, and $n\_points\_per\_class = 1000$ for all $n\_classes$. For small-size datasets, eight datasets are created with $n\_classes =\{5, 10, 15, 20, 25, 30, 35, 40 \}$, $n\_feature= 100$, and $n\_points\_per\_class = 100$.
}

\subsection{Hyper-parameters Setting}
The tested representation learning methods include TAE, MVAE and MAE~\cite{LyVu}, CSAEC \cite{Luo2018CSAEC} \cite{park2020CNN1D}. In addition, we also compare TAE with popular machine learning models including Multi-Layer Perceptron (MLP) \cite{cherfi2025mlp, yin2023igrf}, Support Vector Machine (SVM), Decision Tree (DT), Random Forest (RF), One-dimensional Convolutional Neural Network (1D-CNN)~\cite{1D-CNN}, Multi-Layer Perceptron (MLP) and the state-of-the-art Xgboost (Xgb)~\cite{xgboost-2023, IoTJ-Xgb}. 

The experiments are implemented using two frameworks: TensorFlow and Scikit-learn. All datasets are normalised using Min-Max normalisation. We also use grid search to tune the hyper-parameters for each machine learning model, as shown in Table \ref{tab:grid_search_params}. For the representation learning methods, the final representation is extracted and used as the input to the Decision Tree (DT) model, and the performance of DT is reported in the experiments. For SDSSL~\cite{SDSSL}, only the k-nearest neighbor (KNN) classifier is applied, following its original implementation.

\begin{table*}[!htbp]
	\caption{\normalsize {\color{black} Performance comparison of TAE with Transformers (MLP, 1DCNN), self-distillation (SDSSL, ATAS, COSMOS), and other representation learning methods (CSAEC, MVAE, MAE, and CTVAE).}}
	\label{tab:experimental_results_tae_vs_kd_rl}
	\setlength\tabcolsep{3pt}
	\centering
	\footnotesize	
	
	\begin{tabular}{|c|c|ccccccc|cc|c|ccc|c|}
		\hline
		&                                    & \multicolumn{7}{c|}{\textbf{IoT IDS}}                                                                                                                                                                                                                                                                                                                                                                             & \multicolumn{2}{c|}{\textbf{\begin{tabular}[c]{@{}c@{}}Network\\  IDS\end{tabular}}}                 & \textbf{\begin{tabular}[c]{@{}c@{}}Cloud\\  IDS\end{tabular}} & \multicolumn{3}{c|}{\textbf{Malware}}                                                                                                                     &                                        \\ \cline{3-15}
		\multirow{-2}{*}{\textbf{Scores}} & \multirow{-2}{*}{\textbf{Method}} & \multicolumn{1}{c|}{\textbf{IoT1}}                 & \multicolumn{1}{c|}{\textbf{IoT2}}                          & \multicolumn{1}{c|}{\textbf{IoT3}}                          & \multicolumn{1}{c|}{\textbf{IoT4}}                          & \multicolumn{1}{c|}{\textbf{IoT5}}                          & \multicolumn{1}{c|}{\textbf{IoT6}}                          & \textbf{IoT7}                          & \multicolumn{1}{c|}{\textbf{NSL}}                           & \textbf{UNSW}                          & \textbf{Cloud}                                                & \multicolumn{1}{c|}{\textbf{Mal-1}}                         & \multicolumn{1}{c|}{\textbf{Mal-2}}                & \textbf{Mal-3}                         & \multirow{-2}{*}{\textbf{Avg}}         \\ \hline
		& Transformer(MLP)                   & \multicolumn{1}{c|}{0.672}                         & \multicolumn{1}{c|}{0.626}                                  & \multicolumn{1}{c|}{0.758}                                  & \multicolumn{1}{c|}{0.659}                                  & \multicolumn{1}{c|}{0.725}                                  & \multicolumn{1}{c|}{0.850}                                  & 0.860                                  & \multicolumn{1}{c|}{0.862}                                  & 0.736                                  & 0.976                                                         & \multicolumn{1}{c|}{0.512}                                  & \multicolumn{1}{c|}{0.720}                         & \textbf{0.955}                         & 0.762                                  \\ \cline{2-16} 
		& Transformer(1DCNN)                 & \multicolumn{1}{c|}{0.618}                         & \multicolumn{1}{c|}{0.615}                                  & \multicolumn{1}{c|}{0.755}                                  & \multicolumn{1}{c|}{0.646}                                  & \multicolumn{1}{c|}{0.667}                                  & \multicolumn{1}{c|}{0.813}                                  & 0.844                                  & \multicolumn{1}{c|}{0.861}                                  & \textbf{0.748}                         & 0.887                                                         & \multicolumn{1}{c|}{0.414}                                  & \multicolumn{1}{c|}{0.719}                         & \textbf{0.955}                         & 0.734                                  \\ \cline{2-16} 
		& SDSSL                              & \multicolumn{1}{c|}{\textbf{0.938}}                & \multicolumn{1}{c|}{0.915}                                  & \multicolumn{1}{c|}{0.958}                                  & \multicolumn{1}{c|}{0.945}                                  & \multicolumn{1}{c|}{0.939}                                  & \multicolumn{1}{c|}{0.716}                                  & 0.857                                  & \multicolumn{1}{c|}{0.832}                                  & 0.671                                  & 0.973                                                         & \multicolumn{1}{c|}{0.513}                                  & \multicolumn{1}{c|}{0.901}                         & \textbf{0.955}                         & 0.855                                  \\ \cline{2-16} 
		& ATAS                               & \multicolumn{1}{c|}{0.924}                         & \multicolumn{1}{c|}{0.918}                                  & \multicolumn{1}{c|}{0.941}                                  & \multicolumn{1}{c|}{0.913}                                  & \multicolumn{1}{c|}{0.933}                                  & \multicolumn{1}{c|}{0.746}                                  & 0.878                                  & \multicolumn{1}{c|}{0.819}                                  & 0.647                                  & 0.943                                                         & \multicolumn{1}{c|}{0.512}                                  & \multicolumn{1}{c|}{0.895}                         & \textbf{0.955}                         & 0.848                                  \\ \cline{2-16} 
		& COSMOS                             & \multicolumn{1}{c|}{0.928}                         & \multicolumn{1}{c|}{0.905}                                  & \multicolumn{1}{c|}{0.948}                                  & \multicolumn{1}{c|}{0.936}                                  & \multicolumn{1}{c|}{0.921}                                  & \multicolumn{1}{c|}{0.722}                                  & 0.880                                  & \multicolumn{1}{c|}{0.844}                                  & 0.667                                  & 0.963                                                         & \multicolumn{1}{c|}{0.415}                                  & \multicolumn{1}{c|}{0.720}                         & \textbf{0.955}                         & 0.831                                  \\ \cline{2-16} 
		& SCAEC                              & \multicolumn{1}{c|}{0.923}                         & \multicolumn{1}{c|}{0.914}                                  & \multicolumn{1}{c|}{0.941}                                  & \multicolumn{1}{c|}{0.933}                                  & \multicolumn{1}{c|}{0.928}                                  & \multicolumn{1}{c|}{0.882}                                  & 0.929                                  & \multicolumn{1}{c|}{0.843}                                  & 0.734                                  & 0.980                                                         & \multicolumn{1}{c|}{0.512}                                  & \multicolumn{1}{c|}{0.889}                         & 0.920                                  & 0.872                                  \\ \cline{2-16} 
		& MVAE                               & \multicolumn{1}{c|}{0.688}                         & \multicolumn{1}{c|}{0.608}                                  & \multicolumn{1}{c|}{0.751}                                  & \multicolumn{1}{c|}{0.654}                                  & \multicolumn{1}{c|}{0.713}                                  & \multicolumn{1}{c|}{0.536}                                  & 0.747                                  & \multicolumn{1}{c|}{0.829}                                  & 0.681                                  & 0.867                                                         & \multicolumn{1}{c|}{0.514}                                  & \multicolumn{1}{c|}{0.687}                         & 0.929                                  & 0.708                                  \\ \cline{2-16} 
		& MAE                                & \multicolumn{1}{c|}{0.925}                         & \multicolumn{1}{c|}{0.913}                                  & \multicolumn{1}{c|}{0.950}                                  & \multicolumn{1}{c|}{0.935}                                  & \multicolumn{1}{c|}{0.931}                                  & \multicolumn{1}{c|}{0.876}                                  & 0.933                                  & \multicolumn{1}{c|}{0.842}                                  & 0.704                                  & 0.960                                                         & \multicolumn{1}{c|}{0.512}                                  & \multicolumn{1}{c|}{\textbf{0.917}}                & 0.921                                  & 0.871                                  \\ \cline{2-16} 
		& CTVAE                              & \multicolumn{1}{c|}{0.933}                         & \multicolumn{1}{c|}{0.930}                                  & \multicolumn{1}{c|}{0.965}                                  & \multicolumn{1}{c|}{0.939}                                  & \multicolumn{1}{c|}{0.948}                                  & \multicolumn{1}{c|}{0.931}                                  & 0.934                                  & \multicolumn{1}{c|}{0.863}                                  & 0.741                                  & 0.980                                                         & \multicolumn{1}{c|}{0.513}                                  & \multicolumn{1}{c|}{0.877}                         & 0.921                                  & 0.883                                  \\ \cline{2-16} 
		\multirow{-10}{*}{Accuracy}       & \cellcolor[HTML]{B2B2B2}TAE        & \multicolumn{1}{c|}{\cellcolor[HTML]{B2B2B2}0.931} & \multicolumn{1}{c|}{\cellcolor[HTML]{B2B2B2}\textbf{0.933}} & \multicolumn{1}{c|}{\cellcolor[HTML]{B2B2B2}\textbf{0.985}} & \multicolumn{1}{c|}{\cellcolor[HTML]{B2B2B2}\textbf{0.949}} & \multicolumn{1}{c|}{\cellcolor[HTML]{B2B2B2}\textbf{0.993}} & \multicolumn{1}{c|}{\cellcolor[HTML]{B2B2B2}\textbf{0.959}} & \cellcolor[HTML]{B2B2B2}\textbf{0.975} & \multicolumn{1}{c|}{\cellcolor[HTML]{B2B2B2}\textbf{0.873}} & \cellcolor[HTML]{B2B2B2}0.747          & \cellcolor[HTML]{B2B2B2}\textbf{0.987}                        & \multicolumn{1}{c|}{\cellcolor[HTML]{B2B2B2}\textbf{0.517}} & \multicolumn{1}{c|}{\cellcolor[HTML]{B2B2B2}0.914} & \cellcolor[HTML]{B2B2B2}\textbf{0.955} & \cellcolor[HTML]{B2B2B2}\textbf{0.901} \\ \hline
		& Transformer(MLP)                   & \multicolumn{1}{c|}{0.599}                         & \multicolumn{1}{c|}{0.624}                                  & \multicolumn{1}{c|}{0.616}                                  & \multicolumn{1}{c|}{0.597}                                  & \multicolumn{1}{c|}{0.650}                                  & \multicolumn{1}{c|}{0.778}                                  & 0.798                                  & \multicolumn{1}{c|}{0.645}                                  & 0.436                                  & 0.843                                                         & \multicolumn{1}{c|}{\textbf{0.462}}                         & \multicolumn{1}{c|}{0.634}                         & 0.902                                  & 0.660                                  \\ \cline{2-16} 
		& Transformer(1DCNN)                 & \multicolumn{1}{c|}{0.562}                         & \multicolumn{1}{c|}{0.603}                                  & \multicolumn{1}{c|}{0.598}                                  & \multicolumn{1}{c|}{0.576}                                  & \multicolumn{1}{c|}{0.574}                                  & \multicolumn{1}{c|}{0.755}                                  & 0.776                                  & \multicolumn{1}{c|}{0.620}                                  & 0.449                                  & 0.699                                                         & \multicolumn{1}{c|}{0.299}                                  & \multicolumn{1}{c|}{0.633}                         & 0.902                                  & 0.619                                  \\ \cline{2-16} 
		& SDSSL                              & \multicolumn{1}{c|}{\textbf{0.929}}                & \multicolumn{1}{c|}{0.900}                                  & \multicolumn{1}{c|}{0.951}                                  & \multicolumn{1}{c|}{0.938}                                  & \multicolumn{1}{c|}{0.925}                                  & \multicolumn{1}{c|}{0.706}                                  & 0.842                                  & \multicolumn{1}{c|}{0.793}                                  & 0.633                                  & 0.973                                                         & \multicolumn{1}{c|}{0.433}                                  & \multicolumn{1}{c|}{0.899}                         & 0.957                                  & 0.837                                  \\ \cline{2-16} 
		& ATAS                               & \multicolumn{1}{c|}{0.904}                         & \multicolumn{1}{c|}{0.902}                                  & \multicolumn{1}{c|}{0.919}                                  & \multicolumn{1}{c|}{0.891}                                  & \multicolumn{1}{c|}{0.916}                                  & \multicolumn{1}{c|}{0.726}                                  & 0.870                                  & \multicolumn{1}{c|}{0.781}                                  & 0.612                                  & 0.943                                                         & \multicolumn{1}{c|}{0.432}                                  & \multicolumn{1}{c|}{0.892}                         & 0.957                                  & 0.827                                  \\ \cline{2-16} 
		& COSMOS                             & \multicolumn{1}{c|}{0.909}                         & \multicolumn{1}{c|}{0.878}                                  & \multicolumn{1}{c|}{0.938}                                  & \multicolumn{1}{c|}{0.921}                                  & \multicolumn{1}{c|}{0.903}                                  & \multicolumn{1}{c|}{0.706}                                  & 0.862                                  & \multicolumn{1}{c|}{0.803}                                  & 0.637                                  & 0.962                                                         & \multicolumn{1}{c|}{0.250}                                  & \multicolumn{1}{c|}{0.672}                         & 0.957                                  & 0.800                                  \\ \cline{2-16} 
		& SCAEC                              & \multicolumn{1}{c|}{0.895}                         & \multicolumn{1}{c|}{0.887}                                  & \multicolumn{1}{c|}{0.919}                                  & \multicolumn{1}{c|}{0.915}                                  & \multicolumn{1}{c|}{0.899}                                  & \multicolumn{1}{c|}{0.886}                                  & 0.923                                  & \multicolumn{1}{c|}{0.798}                                  & 0.692                                  & 0.980                                                         & \multicolumn{1}{c|}{0.432}                                  & \multicolumn{1}{c|}{0.887}                         & 0.928                                  & 0.849                                  \\ \cline{2-16} 
		& MVAE                               & \multicolumn{1}{c|}{0.642}                         & \multicolumn{1}{c|}{0.528}                                  & \multicolumn{1}{c|}{0.666}                                  & \multicolumn{1}{c|}{0.536}                                  & \multicolumn{1}{c|}{0.702}                                  & \multicolumn{1}{c|}{0.617}                                  & 0.713                                  & \multicolumn{1}{c|}{0.779}                                  & 0.623                                  & 0.819                                                         & \multicolumn{1}{c|}{0.435}                                  & \multicolumn{1}{c|}{0.683}                         & 0.935                                  & 0.668                                  \\ \cline{2-16} 
		& MAE                                & \multicolumn{1}{c|}{0.900}                         & \multicolumn{1}{c|}{0.885}                                  & \multicolumn{1}{c|}{0.937}                                  & \multicolumn{1}{c|}{0.918}                                  & \multicolumn{1}{c|}{0.908}                                  & \multicolumn{1}{c|}{0.867}                                  & 0.934                                  & \multicolumn{1}{c|}{0.804}                                  & 0.659                                  & 0.960                                                         & \multicolumn{1}{c|}{0.432}                                  & \multicolumn{1}{c|}{\textbf{0.917}}                & 0.929                                  & 0.850                                  \\ \cline{2-16} 
		& CTVAE                              & \multicolumn{1}{c|}{0.915}                         & \multicolumn{1}{c|}{0.916}                                  & \multicolumn{1}{c|}{0.960}                                  & \multicolumn{1}{c|}{0.931}                                  & \multicolumn{1}{c|}{0.937}                                  & \multicolumn{1}{c|}{0.927}                                  & 0.923                                  & \multicolumn{1}{c|}{0.826}                                  & 0.673                                  & 0.979                                                         & \multicolumn{1}{c|}{0.433}                                  & \multicolumn{1}{c|}{0.873}                         & 0.929                                  & 0.863                                  \\ \cline{2-16} 
		\multirow{-10}{*}{F-score}        & \cellcolor[HTML]{B2B2B2}TAE        & \multicolumn{1}{c|}{\cellcolor[HTML]{B2B2B2}0.916} & \multicolumn{1}{c|}{\cellcolor[HTML]{B2B2B2}\textbf{0.922}} & \multicolumn{1}{c|}{\cellcolor[HTML]{B2B2B2}\textbf{0.985}} & \multicolumn{1}{c|}{\cellcolor[HTML]{B2B2B2}\textbf{0.944}} & \multicolumn{1}{c|}{\cellcolor[HTML]{B2B2B2}\textbf{0.993}} & \multicolumn{1}{c|}{\cellcolor[HTML]{B2B2B2}\textbf{0.949}} & \cellcolor[HTML]{B2B2B2}\textbf{0.970} & \multicolumn{1}{c|}{\cellcolor[HTML]{B2B2B2}\textbf{0.844}} & \cellcolor[HTML]{B2B2B2}\textbf{0.695} & \cellcolor[HTML]{B2B2B2}\textbf{0.987}                        & \multicolumn{1}{c|}{\cellcolor[HTML]{B2B2B2}0.439}          & \multicolumn{1}{c|}{\cellcolor[HTML]{B2B2B2}0.912} & \cellcolor[HTML]{B2B2B2}\textbf{0.958} & \cellcolor[HTML]{B2B2B2}\textbf{0.886} \\ \hline
	\end{tabular}

\end{table*}

For three neural network models, i.e., MAE, MVAE, and TAE, only one hidden layer with ReLU activation is used in their subnetworks (the encoder, hermaphrodite, and decoder). The  ADAM optimisation algorithm~\cite{kingma2015adam} is used to train these models. The learning rate $\alpha$ is set at $10^{-4}$, the number of epochs is 5000, and the batch size is set at 100. We also use the validation set (30\% of the training set) to early stop during the training process. If the difference in the loss function between 10 consecutive epochs is lower than a $threshold = 1$, we stop the training process. 
MAE and MVAE models are only investigated on binary datasets in~\cite{LyVu}. In this paper, we extend them for multi-class problems by setting the value of $\mu^{(c)}_{j}$ from $ 2 $ to $ n +2 $, where $n$ is the number of classes. {\color{black} For CSAE, the same architecture as in~\cite{park2020CNN1D} is implemented using 1D convolution. After training CSAE in an unsupervised manner, a Softmax layer is added to the encoder and CSAE is further trained in a supervised manner. 
After supervised training, the Softmax layer is removed and the resulting encoder, referred to as CSAEC, is used to extract the learned representations for downstream evaluation. The CSAEC representations are also evaluated using the same DT classifier.
}
Finally, for the TAE model, we also tune two hyperparameters, i.e. the scale of the transformation operator $S$ in Algorithm \ref{algo:separated_multi_cate} and the dimensionality of the latent space $D_z$. 
The number of neurons in the encoder of the AE-based models is designed as follows: $\{d, 0.9d, 0.8d, 0.7d, d_z \}$, where $d$ is the dimension of the input and $d_z = [\sqrt{d}] + 1$ is the dimension of the latent space \cite{LyVu, CTVAE}. The structure of the decoder is symmetric with the encoder.

For transformer-based models (MLP~\cite{Transformer-MLP} and 1DCNN~\cite{Transforer-1DCNN}), the number of attention heads was tuned. For self-distillation (SD) methods, SDSSL~\cite{SDSSL} was tuned with respect to the latent space dimension and the strategies used to generate different views, including noise standard deviation and shifting operations. For ATAS~\cite{ATAS} and COSMOS~\cite{COSMOS}, the trade-off coefficients among the loss components were optimized. Since SDSSL, ATAS, and COSMOS were originally designed for image domains, we adapted them to IoT cybersecurity datasets using a 1DCNN backbone combined with an attention mechanism employing four attention heads.

{\color{black}
To set up the experiment of fixed codewords \cite{evron2023role}, we generate codewords with a length equal to the dimensionality of the latent space, i.e., $d_z$. Specifically, $d_z$ is tuned from the list $\{d_z, 1.5 \times d_z, 2 \times d_z, 3 \times d_z \}$, where $d_z = [\sqrt{d}] + 1$, and $d$ is the dimensionality of the input data. For MLP, we do not use a softmax layer; instead, we project the data into the latent space using codewords. Note that traditional methods use a softmax layer with a one-hot encoding vector, where the size is equal to the number of classes. For MAE, the bottleneck layer is used as the latent space \cite{LyVu}. The hyper-parameters are set up as observed in Table \ref{tab:grid_search_params}.
}

{\color{black}
	For KD, we consider two teacher architectures, namely DeepMLP and ResNet. A compact MLP is used as the client model for a fair comparison with the proposed TAE. The DeepMLP teacher comprises three fully connected layers with 128, 128, and 64 hidden units, followed by two residual fully connected blocks and a 64-dimensional embedding. The ResNet teacher employs a 128-dimensional fully connected backbone followed by 12 residual blocks and a 128-dimensional embedding. The client MLP uses a compact 64-64-32 backbone followed by two residual blocks and a 32-dimensional embedding, with KD performed using a weighted combination of hard-label and temperature-scaled soft-label losses~\cite{hinton2015distilling}. Since IoT1--IoT5 share the same class definitions, we train the KD teacher models on IoT1 and subsequently train and evaluate the compact student models on IoT2, IoT3, IoT4, and IoT5.
}

\begin{table*}[t]
	\caption{\normalsize Performance of TAE compared to popular machine learning models.}
	\label{tab:experimental_results_03}
	\setlength\tabcolsep{5pt}
	\centering
	\footnotesize	
\begin{tabular}{|c|c|ccccccc|cc|c|ccc|c|}
\hline
&& \multicolumn{7}{c|}{\textbf{IoT IDS}}  & \multicolumn{2}{c|}{\textbf{\begin{tabular}[c]{@{}c@{}}Network\\  IDS\end{tabular}}}& \textbf{\begin{tabular}[c]{@{}c@{}}Cloud\\  IDS\end{tabular}} & \multicolumn{3}{c|}{\textbf{Malware}}  &\\ \cline{3-15}
\multirow{-2}{*}{\textbf{Metric}} & \multirow{-2}{*}{\textbf{Method}} & \multicolumn{1}{c|}{\textbf{IoT1}}  & \multicolumn{1}{c|}{\textbf{IoT2}}  & \multicolumn{1}{c|}{\textbf{IoT3}}  & \multicolumn{1}{c|}{\textbf{IoT4}}  & \multicolumn{1}{c|}{\textbf{IoT5}}  & \multicolumn{1}{c|}{\textbf{IoT6}}  & \textbf{IoT7}  & \multicolumn{1}{c|}{\textbf{NSL}}   & \textbf{UNSW} & \textbf{Cloud}& \multicolumn{1}{c|}{\textbf{Mal-1}} & \multicolumn{1}{c|}{\textbf{Mal-2}} & \textbf{Mal-3} & \multirow{-2}{*}{\textbf{Avg}} \\ \hline \hline
& Xgb& \multicolumn{1}{c|}{0.923}  & \multicolumn{1}{c|}{0.910}  & \multicolumn{1}{c|}{0.941}  & \multicolumn{1}{c|}{0.921}  & \multicolumn{1}{c|}{0.928}  & \multicolumn{1}{c|}{0.941}  & 0.959  & \multicolumn{1}{c|}{0.872}  & \textbf{0.762}& 0.985 & \multicolumn{1}{c|}{0.419}  & \multicolumn{1}{c|}{0.607}  & 0.921  & 0.853  \\ \cline{2-16} 
& SVM& \multicolumn{1}{c|}{0.670}  & \multicolumn{1}{c|}{0.612}  & \multicolumn{1}{c|}{0.751}  & \multicolumn{1}{c|}{0.655}  & \multicolumn{1}{c|}{0.708}  & \multicolumn{1}{c|}{0.837}  & 0.856  & \multicolumn{1}{c|}{0.844}  & 0.672 & 0.890 & \multicolumn{1}{c|}{0.409}  & \multicolumn{1}{c|}{0.720}  & 0.939  & 0.736  \\ \cline{2-16} 
& DT & \multicolumn{1}{c|}{0.672}  & \multicolumn{1}{c|}{0.654}  & \multicolumn{1}{c|}{0.753}  & \multicolumn{1}{c|}{0.656}  & \multicolumn{1}{c|}{0.740}  & \multicolumn{1}{c|}{0.844}  & 0.867  & \multicolumn{1}{c|}{0.866}  & 0.743 & 0.977 & \multicolumn{1}{c|}{0.414}  & \multicolumn{1}{c|}{0.544}  & 0.918  & 0.742  \\ \cline{2-16} 
& RF & \multicolumn{1}{c|}{0.672}  & \multicolumn{1}{c|}{0.613}  & \multicolumn{1}{c|}{0.752}  & \multicolumn{1}{c|}{0.656}  & \multicolumn{1}{c|}{0.709}  & \multicolumn{1}{c|}{0.850}  & 0.858  & \multicolumn{1}{c|}{0.871}  & 0.757 & 0.983 & \multicolumn{1}{c|}{0.416}  & \multicolumn{1}{c|}{0.777}  & 0.921  & 0.757  \\ \cline{2-16} 
& 1D-CNN & \multicolumn{1}{c|}{0.635}  & \multicolumn{1}{c|}{0.613}  & \multicolumn{1}{c|}{0.752}  & \multicolumn{1}{c|}{0.654}  & \multicolumn{1}{c|}{0.709}  & \multicolumn{1}{c|}{0.820}  & 0.846  & \multicolumn{1}{c|}{0.839}  & 0.703 & 0.980 & \multicolumn{1}{c|}{0.474}  & \multicolumn{1}{c|}{0.674}  & 0.911  & 0.739  \\ \cline{2-16} 
& MLP& \multicolumn{1}{c|}{0.706}  & \multicolumn{1}{c|}{0.739}  & \multicolumn{1}{c|}{0.810}  & \multicolumn{1}{c|}{0.733}  & \multicolumn{1}{c|}{0.780}  & \multicolumn{1}{c|}{0.724}  & 0.856  & \multicolumn{1}{c|}{0.861}  & 0.722 & 0.926 & \multicolumn{1}{c|}{0.414}  & \multicolumn{1}{c|}{0.719}  & 0.954  & 0.765  \\ \cline{2-16} 
\multirow{-7}{*}{Accuracy}  & \cellcolor[HTML]{DDDDDD} {{TAE-DT}}& \multicolumn{1}{c|}{\cellcolor[HTML]{DDDDDD}\textbf{0.931}} & \multicolumn{1}{c|}{\cellcolor[HTML]{DDDDDD}\textbf{0.933}} & \multicolumn{1}{c|}{\cellcolor[HTML]{DDDDDD}\textbf{0.985}} & \multicolumn{1}{c|}{\cellcolor[HTML]{DDDDDD}\textbf{0.949}} & \multicolumn{1}{c|}{\cellcolor[HTML]{DDDDDD}\textbf{0.993}} & \multicolumn{1}{c|}{\cellcolor[HTML]{DDDDDD}\textbf{0.959}} & \cellcolor[HTML]{DDDDDD}\textbf{0.975} & \multicolumn{1}{c|}{\cellcolor[HTML]{DDDDDD}\textbf{0.873}} & \cellcolor[HTML]{DDDDDD}0.747 & \cellcolor[HTML]{DDDDDD}\textbf{0.987}& \multicolumn{1}{c|}{\cellcolor[HTML]{DDDDDD}\textbf{0.517}} & \multicolumn{1}{c|}{\cellcolor[HTML]{DDDDDD}\textbf{0.914}} & \cellcolor[HTML]{DDDDDD}\textbf{0.955} & \cellcolor[HTML]{DDDDDD}\textbf{0.901} \\ \hline \hline
& Xgb& \multicolumn{1}{c|}{0.894}  & \multicolumn{1}{c|}{0.879}  & \multicolumn{1}{c|}{0.918}  & \multicolumn{1}{c|}{0.891}  & \multicolumn{1}{c|}{0.900}  & \multicolumn{1}{c|}{0.929}  & 0.945  & \multicolumn{1}{c|}{0.830}  & \textbf{0.733}& 0.985 & \multicolumn{1}{c|}{0.259}  & \multicolumn{1}{c|}{0.608}  & 0.929  & 0.823  \\ \cline{2-16} 
& SVM& \multicolumn{1}{c|}{0.553}  & \multicolumn{1}{c|}{0.475}  & \multicolumn{1}{c|}{0.662}  & \multicolumn{1}{c|}{0.535}  & \multicolumn{1}{c|}{0.603}  & \multicolumn{1}{c|}{0.784}  & 0.805  & \multicolumn{1}{c|}{0.794}  & 0.606 & 0.842 & \multicolumn{1}{c|}{0.238}  & \multicolumn{1}{c|}{0.670}  & 0.944  & 0.655  \\ \cline{2-16} 
& DT & \multicolumn{1}{c|}{0.557}  & \multicolumn{1}{c|}{0.547}  & \multicolumn{1}{c|}{0.665}  & \multicolumn{1}{c|}{0.536}  & \multicolumn{1}{c|}{0.658}  & \multicolumn{1}{c|}{0.799}  & 0.824  & \multicolumn{1}{c|}{0.828}  & 0.720 & 0.977 & \multicolumn{1}{c|}{0.250}  & \multicolumn{1}{c|}{0.545}  & 0.926  & 0.679  \\ \cline{2-16} 
& RF & \multicolumn{1}{c|}{0.557}  & \multicolumn{1}{c|}{0.477}  & \multicolumn{1}{c|}{0.664}  & \multicolumn{1}{c|}{0.536}  & \multicolumn{1}{c|}{0.604}  & \multicolumn{1}{c|}{0.797}  & 0.807  & \multicolumn{1}{c|}{0.827}  & 0.727 & 0.982 & \multicolumn{1}{c|}{0.252}  & \multicolumn{1}{c|}{0.751}  & 0.929  & 0.685  \\ \cline{2-16} 
& 1D-CNN & \multicolumn{1}{c|}{0.515}  & \multicolumn{1}{c|}{0.477}  & \multicolumn{1}{c|}{0.663}  & \multicolumn{1}{c|}{0.534}  & \multicolumn{1}{c|}{0.605}  & \multicolumn{1}{c|}{0.768}  & 0.796  & \multicolumn{1}{c|}{0.788}  & 0.646 & 0.979 & \multicolumn{1}{c|}{0.366}  & \multicolumn{1}{c|}{0.635}  & 0.920  & 0.669  \\ \cline{2-16} 
& MLP& \multicolumn{1}{c|}{0.605}  & \multicolumn{1}{c|}{0.651}  & \multicolumn{1}{c|}{0.744}  & \multicolumn{1}{c|}{0.645}  & \multicolumn{1}{c|}{0.703}  & \multicolumn{1}{c|}{0.635}  & 0.815  & \multicolumn{1}{c|}{0.822}  & 0.671 & 0.913 & \multicolumn{1}{c|}{0.248}  & \multicolumn{1}{c|}{0.670}  & 0.956  & 0.698  \\ \cline{2-16} 
\multirow{-7}{*}{F1-score}   & \cellcolor[HTML]{DDDDDD}TAE-DT& \multicolumn{1}{c|}{\cellcolor[HTML]{DDDDDD}\textbf{0.916}} & \multicolumn{1}{c|}{\cellcolor[HTML]{DDDDDD}\textbf{0.922}} & \multicolumn{1}{c|}{\cellcolor[HTML]{DDDDDD}\textbf{0.985}} & \multicolumn{1}{c|}{\cellcolor[HTML]{DDDDDD}\textbf{0.944}} & \multicolumn{1}{c|}{\cellcolor[HTML]{DDDDDD}\textbf{0.993}} & \multicolumn{1}{c|}{\cellcolor[HTML]{DDDDDD}\textbf{0.949}} & \cellcolor[HTML]{DDDDDD}\textbf{0.970} & \multicolumn{1}{c|}{\cellcolor[HTML]{DDDDDD}\textbf{0.844}} & \cellcolor[HTML]{DDDDDD}0.695 & \cellcolor[HTML]{DDDDDD}\textbf{0.987}& \multicolumn{1}{c|}{\cellcolor[HTML]{DDDDDD}\textbf{0.439}} & \multicolumn{1}{c|}{\cellcolor[HTML]{DDDDDD}\textbf{0.912}} & \cellcolor[HTML]{DDDDDD}\textbf{0.958} & \cellcolor[HTML]{DDDDDD}\textbf{0.886} \\ \hline 

\end{tabular}

\end{table*}

\section{PERFORMANCE ANALYSIS}
\label{sec:performance_analysis}
\subsection{Accuracy of Detection Models}

{\color{black}
	Table~\ref{tab:experimental_results_tae_vs_kd_rl} compares TAE with transformer-based models (MLP and 1DCNN), self-distillation methods (SDSSL, ATAS, and COSMOS), and other representation learning (RL) methods (CSAEC, MVAE, MAE, and CTVAE) across 13 IoT datasets. TAE consistently achieves the highest accuracy and F1-score, with an average accuracy of $0.901$, outperforming MLP ($0.762$), 1DCNN ($0.734$), SDSSL ($0.802$), ATAS ($0.848$), COSMOS ($0.831$), CSAEC ($0.872$), MVAE ($0.708$), MAE ($0.871$), and CTVAE ($0.883$). The performance advantage of TAE is attributed to its deterministic class-adaptive representation, which is better suited to the heterogeneous feature structures of IoT data than image-oriented self-distillation methods and codeword-based representation learning approaches. In particular, deterministic transformation avoids posterior collapse encountered by VAE-based methods such as MVAE and CTVAE.
}

Table \ref{tab:experimental_results_03} presents the performance of TAE compared to six popular machine learning models: SVM, DT, RF, 1D-CNN, MLP, and Xgb, across the tested datasets. Two notable results emerge from this table. First, TAE consistently outperforms the other models, except Xgb. For instance, on the IoT1 dataset, TAE achieves an accuracy of 0.931, while the accuracies of SVM, DT, RF, 1D-CNN, and MLP are 0.670, 0.672, 0.672, 0.635, and 0.706, respectively. This suggests that the original data representations in the tested problems pose challenges for conventional machine learning methods. However, by transforming the original representation into a more separable form and subsequently reconstructing it through TAE's reconstruction mechanism, the data becomes more amenable to machine learning models.
Second, the model closest in performance to TAE is Xgb. Despite this, TAE still outperforms Xgb, demonstrating its superior capability in cyberattack detection. As a result, TAE proves to be more effective and applicable in this domain than Xgb.


Overall, the above results show the superior performance of the proposed model, i.e., TAE over the other representation learning and machine learning algorithms on a wide range of cybersecurity problems. TAE is trained in the supervised mode, and its performance is usually better than that of the state-of-the-art representation models, i.e., MAE and SCAEC. TAE also shows superior performance over well-known machine learning models using the original datasets, i.e., SVM, DT, RF, 1D-CNN, and MLP. In addition, the results of TAE are almost higher than the advanced ensemble learning method (Xgb). 

\begin{table}[t]
	\caption{\normalsize Confusion matrix of TAE, MAE, and Xgb on the NSL dataset.}
	\label{tab:confusio_matrix_NSL}
	\centering
	\footnotesize
	\setlength\tabcolsep{3pt}
\begin{tabular}{|c|c|c|c|c|c|c|}
\hline
\textbf{Metthods}& \textbf{Classes} & \textbf{Normal} & \textbf{DoS}  & \textbf{U2R}& \textbf{R2L}& \textbf{Probe} \\ \hline \hline
 & \textbf{Normal}  & \textbf{9475}& 56  & 5 & 23 & 152  \\ \cline{2-7} 
 & \textbf{DoS}& 150& \textbf{5557} & 0 & 0  & 34\\ \cline{2-7} 
 & \textbf{U2R}& 23& 2& \cellcolor[HTML]{B2B2B2}\textbf{10} & 0  & 2\\ \cline{2-7} 
 & \textbf{R2L}& 1689  & 7& 22& \cellcolor[HTML]{B2B2B2}\textbf{475} & 6\\ \cline{2-7} 
\multirow{-5}{*}{\textbf{TAE}} & \textbf{Probe}& 223& 2& 0 & 0  & \textbf{881}\\ \hline \hline
 & \textbf{Normal}  & \textbf{9441}& 93  & 4 & 14 & 159  \\ \cline{2-7} 
 & \textbf{DoS}& 301& \textbf{5333} & 1 & 11 & 95\\ \cline{2-7} 
 & \textbf{U2R}& 25& 7& \cellcolor[HTML]{B2B2B2}\textbf{1}  & 2  & 2\\ \cline{2-7} 
 & \textbf{R2L}& 1986  & 109 & 3 & \cellcolor[HTML]{B2B2B2}\textbf{79}  & 22\\ \cline{2-7} 
\multirow{-5}{*}{\textbf{MAE}} & \textbf{Probe}& 253& 7& 1 & 0  & \textbf{845}\\ \hline \hline
 & \textbf{Normal}  & \textbf{9428}& 81  & 1 & 0  & 201  \\ \cline{2-7} 
 & \textbf{DoS}& 12& \textbf{5716} & 0 & 0  & 13\\ \cline{2-7} 
 & \textbf{U2R}& 31& 0& \cellcolor[HTML]{B2B2B2}\textbf{4}  & 2  & 0\\ \cline{2-7} 
 & \textbf{R2L}& 2037  & 0& 1 & \cellcolor[HTML]{B2B2B2}\textbf{135} & 26\\ \cline{2-7} 
\multirow{-5}{*}{\textbf{Xgb}} & \textbf{Probe}& 1& 0& 0 & 0  & \textbf{1105}  \\ \hline
\end{tabular}
 
\end{table}

\begin{table}[t]
	\caption{\normalsize Confusion matrix of TAE, MAE, and Xgb on the Cloud dataset.}
	\label{tab:confusio_matrix_cloud}
	\centering
	\footnotesize
	\setlength\tabcolsep{3.5pt}
\begin{tabular}{|c|r|c|c|c|c|c|c|c|}
\hline
\textbf{Method}  & \multicolumn{1}{c|}{\textbf{Classes}} & $l_1$ & $l_2$  & $l_3$& $l_4$  & $l_5$ & $l_6$ & $l_7$\\ \hline \hline
 & normal traffic ($l_1$)& \textbf{1151} & 0& 0 & 5& 0 & 0 & 0 \\ \cline{2-9} 
 & Dns-flood ($l_2$)& 0& \textbf{6} & 0 & 0& 0 & 0 & 5 \\ \cline{2-9} 
 & Ping-of-death ($l_3$)& 0& 0& \cellcolor[HTML]{CCCCCC}\textbf{18} & 0& 0 & 0 & 0 \\ \cline{2-9} 
 & Slowloris ($l_4$)& 7& 0& 0 & \cellcolor[HTML]{CCCCCC}\textbf{138} & 0 & 0 & 0 \\ \cline{2-9} 
 & Tcp-land-attack ($l_5$)  & 0& 0& 0 & 0& \cellcolor[HTML]{CCCCCC}\textbf{11} & 0 & 0 \\ \cline{2-9} 
 & TCP-syn-flood ($l_6$)& 0& 0& 0 & 0& 0 & \textbf{19} & 0 \\ \cline{2-9} 
\multirow{-7}{*}{TAE} & Udp-flood ($l_7$)& 0& 1& 0 & 0& 0 & 0 & \textbf{13} \\ \hline \hline
 & normal traffic ($l_1$)& \textbf{1144} & 0& 0 & 12 & 0 & 0 & 0 \\ \cline{2-9} 
 & Dns-flood ($l_2$)& 0& \textbf{5} & 0 & 0& 1 & 1 & 4 \\ \cline{2-9} 
 & Ping-of-death ($l_3$)& 0& 1& \cellcolor[HTML]{CCCCCC}\textbf{12} & 0& 2 & 3 & 0 \\ \cline{2-9} 
 & Slowloris ($l_4$)& 28  & 0& 0 & \cellcolor[HTML]{CCCCCC}\textbf{117} & 0 & 0 & 0 \\ \cline{2-9} 
 & Tcp-land-attack ($l_5$)  & 0& 0& 0 & 0& \cellcolor[HTML]{CCCCCC}\textbf{8}  & 2 & 1 \\ \cline{2-9} 
 & TCP-syn-flood ($l_6$)& 0& 1& 1 & 0& 3 & \textbf{13} & 1 \\ \cline{2-9} 
\multirow{-7}{*}{MAE} & Udp-flood ($l_7$)& 0& 3& 0 & 0& 0 & 2 & \textbf{9}  \\ \hline \hline
 & normal traffic ($l_1$)& \textbf{1153} & 0& 0 & 3& 0 & 0 & 0 \\ \cline{2-9} 
 & Dns-flood ($l_2$)& 0& \textbf{6} & 0 & 0& 0 & 0 & 5 \\ \cline{2-9} 
 & Ping-of-death ($l_3$)& 0& 0&\cellcolor[HTML]{CCCCCC} \textbf{18} & 0& 0 & 0 & 0 \\ \cline{2-9} 
 & Slowloris ($l_4$)& 9& 0& 0 & \cellcolor[HTML]{CCCCCC}\textbf{136} & 0 & 0 & 0 \\ \cline{2-9} 
 & Tcp-land-attack ($l_5$)  & 0& 0& 0 & 0& \cellcolor[HTML]{CCCCCC}\textbf{11} & 0 & 0 \\ \cline{2-9} 
 & Tcp-syn-flood ($l_6$)& 0& 0& 0 & 0& 0 & \textbf{19} & 0 \\ \cline{2-9} 
\multirow{-7}{*}{Xgb} & Udp-flood ($l_7$)& 0& 3& 0 & 0& 0 & 0 & \textbf{11} \\ \hline
\end{tabular}

\end{table}

\begin{table}[htp]
	\caption{\normalsize Confusion matrix of TAE, MAE, and Xgb on the Mal3 dataset.}
	\label{tab:confusio_matrix_mal3}
	\centering
	\footnotesize
	\setlength\tabcolsep{4.5pt}
\begin{tabular}{|c|c|cc|}
\hline
\multirow{2}{*}{\textbf{Method}} & \multirow{2}{*}{\textbf{Classes}} & \multicolumn{2}{c|}{\textbf{Mal3}}\\ \cline{3-4} 
  && \multicolumn{1}{c|}{Normal} & Attacks \\ \hline \hline
\multirow{2}{*}{TAE}& Normal& \multicolumn{1}{c|}{\textbf{5586}} & 344  \\ \cline{2-4} 
  & Attacks& \multicolumn{1}{c|}{2059}& \textbf{45635} \\ \hline \hline
\multirow{2}{*}{MAE}& Normal& \multicolumn{1}{c|}{\textbf{5725}} & 205  \\ \cline{2-4} 
  & Attacks& \multicolumn{1}{c|}{4013}& \textbf{43681} \\ \hline \hline
\multirow{2}{*}{Xgb}& Normal& \multicolumn{1}{c|}{\textbf{5677}} & 253  \\ \cline{2-4} 
  & Attacks& \multicolumn{1}{c|}{3986}& \textbf{43708} \\ \hline
\end{tabular}

\end{table}

\begin{table*}[t]
	\caption{\normalsize {\color{black} Performance of TAE's {\color {black} CATR} compared to codewords for representation learning models.}}
	\label{tab:result_codewords_inblack}
	\setlength\tabcolsep{4pt}
	\centering
	\footnotesize	
 \begin{tabular}{|c|c|c|ccccccc|c|}
\hline
{\color[HTML]{000000} }& {\color[HTML]{000000} }& {\color[HTML]{000000} }& \multicolumn{7}{c|}{{\color[HTML]{000000} \textbf{Datasets}}}& {\color[HTML]{000000} }  \\ \cline{4-10}
\multirow{-2}{*}{{\color[HTML]{000000} \textbf{M}}} & \multirow{-2}{*}{{\color[HTML]{000000} \textbf{\begin{tabular}[c]{@{}c@{}}Representation\\ models\end{tabular}}}} & \multirow{-2}{*}{{\color[HTML]{000000} \textbf{Codewords}}} & \multicolumn{1}{c|}{{\color[HTML]{000000} \textbf{IoT1}}}  & \multicolumn{1}{c|}{{\color[HTML]{000000} \textbf{IoT2}}} & \multicolumn{1}{c|}{{\color[HTML]{000000} \textbf{IoT3}}} & \multicolumn{1}{c|}{{\color[HTML]{000000} \textbf{IoT4}}}  & \multicolumn{1}{c|}{{\color[HTML]{000000} \textbf{IoT5}}} & \multicolumn{1}{c|}{{\color[HTML]{000000} \textbf{IoT6}}} & {\color[HTML]{000000} \textbf{IoT7}}  & \multirow{-2}{*}{{\color[HTML]{000000} \textbf{Avg}}} \\ \hline \hline
{\color[HTML]{000000} }& \cellcolor[HTML]{CCCCCC}{\color[HTML]{000000} }& \cellcolor[HTML]{CCCCCC}{\color[HTML]{000000} One-hot} & \multicolumn{1}{c|}{\cellcolor[HTML]{CCCCCC}{\color[HTML]{000000} 0.923}}  & \multicolumn{1}{c|}{\cellcolor[HTML]{CCCCCC}{\color[HTML]{000000} 0.925}} & \multicolumn{1}{c|}{\cellcolor[HTML]{CCCCCC}{\color[HTML]{000000} 0.957}} & \multicolumn{1}{c|}{\cellcolor[HTML]{CCCCCC}{\color[HTML]{000000} \textbf{0.953}}} & \multicolumn{1}{c|}{\cellcolor[HTML]{CCCCCC}{\color[HTML]{000000} 0.937}} & \multicolumn{1}{c|}{\cellcolor[HTML]{CCCCCC}{\color[HTML]{000000} 0.892}} & \cellcolor[HTML]{CCCCCC}{\color[HTML]{000000} 0.922}  & \cellcolor[HTML]{CCCCCC}{\color[HTML]{000000} 0.930}  \\ \cline{3-11} 
{\color[HTML]{000000} }& \cellcolor[HTML]{CCCCCC}{\color[HTML]{000000} }& \cellcolor[HTML]{CCCCCC}{\color[HTML]{000000} Hadamard}& \multicolumn{1}{c|}{\cellcolor[HTML]{CCCCCC}{\color[HTML]{000000} \textbf{0.939}}} & \multicolumn{1}{c|}{\cellcolor[HTML]{CCCCCC}{\color[HTML]{000000} 0.919}} & \multicolumn{1}{c|}{\cellcolor[HTML]{CCCCCC}{\color[HTML]{000000} 0.955}} & \multicolumn{1}{c|}{\cellcolor[HTML]{CCCCCC}{\color[HTML]{000000} 0.945}}  & \multicolumn{1}{c|}{\cellcolor[HTML]{CCCCCC}{\color[HTML]{000000} 0.950}} & \multicolumn{1}{c|}{\cellcolor[HTML]{CCCCCC}{\color[HTML]{000000} 0.930}} & \cellcolor[HTML]{CCCCCC}{\color[HTML]{000000} 0.938}  & \cellcolor[HTML]{CCCCCC}{\color[HTML]{000000} 0.939}  \\ \cline{3-11} 
{\color[HTML]{000000} }& \multirow{-3}{*}{\cellcolor[HTML]{CCCCCC}{\color[HTML]{000000} MLP}} & \cellcolor[HTML]{CCCCCC}{\color[HTML]{000000} TAE's means} & \multicolumn{1}{c|}{\cellcolor[HTML]{CCCCCC}{\color[HTML]{000000} 0.923}}  & \multicolumn{1}{c|}{\cellcolor[HTML]{CCCCCC}{\color[HTML]{000000} 0.914}} & \multicolumn{1}{c|}{\cellcolor[HTML]{CCCCCC}{\color[HTML]{000000} 0.971}} & \multicolumn{1}{c|}{\cellcolor[HTML]{CCCCCC}{\color[HTML]{000000} 0.925}}  & \multicolumn{1}{c|}{\cellcolor[HTML]{CCCCCC}{\color[HTML]{000000} 0.949}} & \multicolumn{1}{c|}{\cellcolor[HTML]{CCCCCC}{\color[HTML]{000000} 0.949}} & \cellcolor[HTML]{CCCCCC}{\color[HTML]{000000} 0.972}  & \cellcolor[HTML]{CCCCCC}{\color[HTML]{000000} 0.943}  \\ \cline{2-11} 
{\color[HTML]{000000} }& \cellcolor[HTML]{DDE8CB}{\color[HTML]{000000} }& \cellcolor[HTML]{DDE8CB}{\color[HTML]{000000} MAE's codeword} & \multicolumn{1}{c|}{\cellcolor[HTML]{DDE8CB}{\color[HTML]{000000} 0.925}}  & \multicolumn{1}{c|}{\cellcolor[HTML]{DDE8CB}{\color[HTML]{000000} 0.913}} & \multicolumn{1}{c|}{\cellcolor[HTML]{DDE8CB}{\color[HTML]{000000} 0.950}} & \multicolumn{1}{c|}{\cellcolor[HTML]{DDE8CB}{\color[HTML]{000000} 0.935}}  & \multicolumn{1}{c|}{\cellcolor[HTML]{DDE8CB}{\color[HTML]{000000} 0.931}} & \multicolumn{1}{c|}{\cellcolor[HTML]{DDE8CB}{\color[HTML]{000000} 0.876}} & \cellcolor[HTML]{DDE8CB}{\color[HTML]{000000} 0.933}  & \cellcolor[HTML]{DDE8CB}{\color[HTML]{000000} 0.923}  \\ \cline{3-11} 
{\color[HTML]{000000} }& \cellcolor[HTML]{DDE8CB}{\color[HTML]{000000} }& \cellcolor[HTML]{DDE8CB}{\color[HTML]{000000} One-hot} & \multicolumn{1}{c|}{\cellcolor[HTML]{DDE8CB}{\color[HTML]{000000} 0.923}}  & \multicolumn{1}{c|}{\cellcolor[HTML]{DDE8CB}{\color[HTML]{000000} 0.910}} & \multicolumn{1}{c|}{\cellcolor[HTML]{DDE8CB}{\color[HTML]{000000} 0.755}} & \multicolumn{1}{c|}{\cellcolor[HTML]{DDE8CB}{\color[HTML]{000000} 0.921}}  & \multicolumn{1}{c|}{\cellcolor[HTML]{DDE8CB}{\color[HTML]{000000} 0.932}} & \multicolumn{1}{c|}{\cellcolor[HTML]{DDE8CB}{\color[HTML]{000000} 0.892}} & \cellcolor[HTML]{DDE8CB}{\color[HTML]{000000} 0.946}  & \cellcolor[HTML]{DDE8CB}{\color[HTML]{000000} 0.897}  \\ \cline{3-11} 
{\color[HTML]{000000} }& \cellcolor[HTML]{DDE8CB}{\color[HTML]{000000} }& \cellcolor[HTML]{DDE8CB}{\color[HTML]{000000} Hadamard}& \multicolumn{1}{c|}{\cellcolor[HTML]{DDE8CB}{\color[HTML]{000000} 0.923}}  & \multicolumn{1}{c|}{\cellcolor[HTML]{DDE8CB}{\color[HTML]{000000} 0.916}} & \multicolumn{1}{c|}{\cellcolor[HTML]{DDE8CB}{\color[HTML]{000000} 0.945}} & \multicolumn{1}{c|}{\cellcolor[HTML]{DDE8CB}{\color[HTML]{000000} 0.930}}  & \multicolumn{1}{c|}{\cellcolor[HTML]{DDE8CB}{\color[HTML]{000000} 0.931}} & \multicolumn{1}{c|}{\cellcolor[HTML]{DDE8CB}{\color[HTML]{000000} 0.940}} & \cellcolor[HTML]{DDE8CB}{\color[HTML]{000000} 0.962}  & \cellcolor[HTML]{DDE8CB}{\color[HTML]{000000} 0.935}  \\ \cline{3-11} 
{\color[HTML]{000000} }& \multirow{-4}{*}{\cellcolor[HTML]{DDE8CB}{\color[HTML]{000000} MAE}} & \cellcolor[HTML]{DDE8CB}{\color[HTML]{000000} TAE's means} & \multicolumn{1}{c|}{\cellcolor[HTML]{DDE8CB}{\color[HTML]{000000} 0.923}}  & \multicolumn{1}{c|}{\cellcolor[HTML]{DDE8CB}{\color[HTML]{000000} 0.917}} & \multicolumn{1}{c|}{\cellcolor[HTML]{DDE8CB}{\color[HTML]{000000} 0.980}} & \multicolumn{1}{c|}{\cellcolor[HTML]{DDE8CB}{\color[HTML]{000000} 0.940}}  & \multicolumn{1}{c|}{\cellcolor[HTML]{DDE8CB}{\color[HTML]{000000} 0.944}} & \multicolumn{1}{c|}{\cellcolor[HTML]{DDE8CB}{\color[HTML]{000000} 0.928}} & \cellcolor[HTML]{DDE8CB}{\color[HTML]{000000} 0.964}  & \cellcolor[HTML]{DDE8CB}{\color[HTML]{000000} 0.942}  \\ \cline{2-11} 
\multirow{-8}{*}{{\color[HTML]{000000} Acc}}& {\color[HTML]{000000} TAE}& {\color[HTML]{000000} {\color {black} CATR}}& \multicolumn{1}{c|}{{\color[HTML]{000000} 0.931}}  & \multicolumn{1}{c|}{{\color[HTML]{000000} \textbf{0.933}}}& \multicolumn{1}{c|}{{\color[HTML]{000000} \textbf{0.985}}}& \multicolumn{1}{c|}{{\color[HTML]{000000} 0.949}}  & \multicolumn{1}{c|}{{\color[HTML]{000000} \textbf{0.993}}}& \multicolumn{1}{c|}{{\color[HTML]{000000} \textbf{0.959}}}& {\color[HTML]{000000} \textbf{0.975}} & {\color[HTML]{000000} \textbf{0.961}} \\ \hline \hline
{\color[HTML]{000000} }& \cellcolor[HTML]{CCCCCC}{\color[HTML]{000000} }& \cellcolor[HTML]{CCCCCC}{\color[HTML]{000000} One-hot} & \multicolumn{1}{c|}{\cellcolor[HTML]{CCCCCC}{\color[HTML]{000000} 0.896}}  & \multicolumn{1}{c|}{\cellcolor[HTML]{CCCCCC}{\color[HTML]{000000} 0.907}} & \multicolumn{1}{c|}{\cellcolor[HTML]{CCCCCC}{\color[HTML]{000000} 0.948}} & \multicolumn{1}{c|}{\cellcolor[HTML]{CCCCCC}{\color[HTML]{000000} \textbf{0.947}}} & \multicolumn{1}{c|}{\cellcolor[HTML]{CCCCCC}{\color[HTML]{000000} 0.918}} & \multicolumn{1}{c|}{\cellcolor[HTML]{CCCCCC}{\color[HTML]{000000} 0.882}} & \cellcolor[HTML]{CCCCCC}{\color[HTML]{000000} 0.914}  & \cellcolor[HTML]{CCCCCC}{\color[HTML]{000000} 0.916}  \\ \cline{3-11} 
{\color[HTML]{000000} }& \cellcolor[HTML]{CCCCCC}{\color[HTML]{000000} }& \cellcolor[HTML]{CCCCCC}{\color[HTML]{000000} Hadamard}& \multicolumn{1}{c|}{\cellcolor[HTML]{CCCCCC}{\color[HTML]{000000} \textbf{0.925}}} & \multicolumn{1}{c|}{\cellcolor[HTML]{CCCCCC}{\color[HTML]{000000} 0.897}} & \multicolumn{1}{c|}{\cellcolor[HTML]{CCCCCC}{\color[HTML]{000000} 0.945}} & \multicolumn{1}{c|}{\cellcolor[HTML]{CCCCCC}{\color[HTML]{000000} 0.935}}  & \multicolumn{1}{c|}{\cellcolor[HTML]{CCCCCC}{\color[HTML]{000000} 0.941}} & \multicolumn{1}{c|}{\cellcolor[HTML]{CCCCCC}{\color[HTML]{000000} 0.921}} & \cellcolor[HTML]{CCCCCC}{\color[HTML]{000000} 0.924}  & \cellcolor[HTML]{CCCCCC}{\color[HTML]{000000} 0.927}  \\ \cline{3-11} 
{\color[HTML]{000000} }& \multirow{-3}{*}{\cellcolor[HTML]{CCCCCC}{\color[HTML]{000000} MLP}} & \cellcolor[HTML]{CCCCCC}{\color[HTML]{000000} TAE's means} & \multicolumn{1}{c|}{\cellcolor[HTML]{CCCCCC}{\color[HTML]{000000} 0.896}}  & \multicolumn{1}{c|}{\cellcolor[HTML]{CCCCCC}{\color[HTML]{000000} 0.888}} & \multicolumn{1}{c|}{\cellcolor[HTML]{CCCCCC}{\color[HTML]{000000} 0.968}} & \multicolumn{1}{c|}{\cellcolor[HTML]{CCCCCC}{\color[HTML]{000000} 0.899}}  & \multicolumn{1}{c|}{\cellcolor[HTML]{CCCCCC}{\color[HTML]{000000} 0.938}} & \multicolumn{1}{c|}{\cellcolor[HTML]{CCCCCC}{\color[HTML]{000000} 0.940}} & \cellcolor[HTML]{CCCCCC}{\color[HTML]{000000} \textbf{0.972}} & \cellcolor[HTML]{CCCCCC}{\color[HTML]{000000} 0.929}  \\ \cline{2-11} 
{\color[HTML]{000000} }& \cellcolor[HTML]{DDE8CB}{\color[HTML]{000000} }& \cellcolor[HTML]{DDE8CB}{\color[HTML]{000000} MAE's codeword} & \multicolumn{1}{c|}{\cellcolor[HTML]{DDE8CB}{\color[HTML]{000000} 0.900}}  & \multicolumn{1}{c|}{\cellcolor[HTML]{DDE8CB}{\color[HTML]{000000} 0.885}} & \multicolumn{1}{c|}{\cellcolor[HTML]{DDE8CB}{\color[HTML]{000000} 0.937}} & \multicolumn{1}{c|}{\cellcolor[HTML]{DDE8CB}{\color[HTML]{000000} 0.918}}  & \multicolumn{1}{c|}{\cellcolor[HTML]{DDE8CB}{\color[HTML]{000000} 0.908}} & \multicolumn{1}{c|}{\cellcolor[HTML]{DDE8CB}{\color[HTML]{000000} 0.867}} & \cellcolor[HTML]{DDE8CB}{\color[HTML]{000000} 0.934}  & \cellcolor[HTML]{DDE8CB}{\color[HTML]{000000} 0.907}  \\ \cline{3-11} 
{\color[HTML]{000000} }& \cellcolor[HTML]{DDE8CB}{\color[HTML]{000000} }& \cellcolor[HTML]{DDE8CB}{\color[HTML]{000000} One-hot} & \multicolumn{1}{c|}{\cellcolor[HTML]{DDE8CB}{\color[HTML]{000000} 0.896}}  & \multicolumn{1}{c|}{\cellcolor[HTML]{DDE8CB}{\color[HTML]{000000} 0.879}} & \multicolumn{1}{c|}{\cellcolor[HTML]{DDE8CB}{\color[HTML]{000000} 0.670}} & \multicolumn{1}{c|}{\cellcolor[HTML]{DDE8CB}{\color[HTML]{000000} 0.892}}  & \multicolumn{1}{c|}{\cellcolor[HTML]{DDE8CB}{\color[HTML]{000000} 0.908}} & \multicolumn{1}{c|}{\cellcolor[HTML]{DDE8CB}{\color[HTML]{000000} 0.879}} & \cellcolor[HTML]{DDE8CB}{\color[HTML]{000000} 0.938}  & \cellcolor[HTML]{DDE8CB}{\color[HTML]{000000} 0.866}  \\ \cline{3-11} 
{\color[HTML]{000000} }& \cellcolor[HTML]{DDE8CB}{\color[HTML]{000000} }& \cellcolor[HTML]{DDE8CB}{\color[HTML]{000000} Hadamard}& \multicolumn{1}{c|}{\cellcolor[HTML]{DDE8CB}{\color[HTML]{000000} 0.895}}  & \multicolumn{1}{c|}{\cellcolor[HTML]{DDE8CB}{\color[HTML]{000000} 0.892}} & \multicolumn{1}{c|}{\cellcolor[HTML]{DDE8CB}{\color[HTML]{000000} 0.925}} & \multicolumn{1}{c|}{\cellcolor[HTML]{DDE8CB}{\color[HTML]{000000} 0.909}}  & \multicolumn{1}{c|}{\cellcolor[HTML]{DDE8CB}{\color[HTML]{000000} 0.906}} & \multicolumn{1}{c|}{\cellcolor[HTML]{DDE8CB}{\color[HTML]{000000} 0.928}} & \cellcolor[HTML]{DDE8CB}{\color[HTML]{000000} 0.949}  & \cellcolor[HTML]{DDE8CB}{\color[HTML]{000000} 0.915}  \\ \cline{3-11} 
{\color[HTML]{000000} }& \multirow{-4}{*}{\cellcolor[HTML]{DDE8CB}{\color[HTML]{000000} MAE}} & \cellcolor[HTML]{DDE8CB}{\color[HTML]{000000} TAE's means} & \multicolumn{1}{c|}{\cellcolor[HTML]{DDE8CB}{\color[HTML]{000000} 0.895}}  & \multicolumn{1}{c|}{\cellcolor[HTML]{DDE8CB}{\color[HTML]{000000} 0.893}} & \multicolumn{1}{c|}{\cellcolor[HTML]{DDE8CB}{\color[HTML]{000000} 0.979}} & \multicolumn{1}{c|}{\cellcolor[HTML]{DDE8CB}{\color[HTML]{000000} 0.927}}  & \multicolumn{1}{c|}{\cellcolor[HTML]{DDE8CB}{\color[HTML]{000000} 0.932}} & \multicolumn{1}{c|}{\cellcolor[HTML]{DDE8CB}{\color[HTML]{000000} 0.918}} & \cellcolor[HTML]{DDE8CB}{\color[HTML]{000000} 0.959}  & \cellcolor[HTML]{DDE8CB}{\color[HTML]{000000} 0.929}  \\ \cline{2-11} 
\multirow{-8}{*}{{\color[HTML]{000000} F1-score}}& {\color[HTML]{000000} TAE}& {\color[HTML]{000000} {\color {black} CATR}}& \multicolumn{1}{c|}{{\color[HTML]{000000} 0.916}}  & \multicolumn{1}{c|}{{\color[HTML]{000000} \textbf{0.922}}}& \multicolumn{1}{c|}{{\color[HTML]{000000} \textbf{0.985}}}& \multicolumn{1}{c|}{{\color[HTML]{000000} 0.944}}  & \multicolumn{1}{c|}{{\color[HTML]{000000} \textbf{0.993}}}& \multicolumn{1}{c|}{{\color[HTML]{000000} \textbf{0.949}}}& {\color[HTML]{000000} 0.970}  & {\color[HTML]{000000} \textbf{0.954}} \\ \hline
\end{tabular}
 
\end{table*}

\subsection{Detecting Sophisticated Attacks}


R2L and U2R are sophisticated attacks in the NSL dataset that challenge conventional machine learning models~\cite{xiang2008design}. U2R attacks exploit system vulnerabilities to gain superuser or root privileges, resulting in subtle changes in system behavior that can be hard to distinguish from normal activities. R2L attacks occur when an attacker attempts unauthorized access to a remote system, often through vulnerabilities or brute force on login credentials, typically involving low network traffic, making them difficult to detect.
The \textcolor{black}{confusion matrix} of the tested methods on the NSL dataset is shown in Table \ref{tab:confusio_matrix_NSL}. This table shows that TAE detects attacks better than MAE and Xgb. Specifically, TAE detects 10 out of 37 U2R attack samples, while MAE and Xgb detect only 1 and 4, respectively. For R2L attacks, TAE correctly detects 475 out of 2199, while MAE and Xgb detect only 79 and 135, respectively.

Ping-of-death, Slowloris, and TCP-land attacks are three popular network disruption attacks on the Cloud dataset~\cite{CloudIDS}. Ping-of-death attacks exploit vulnerabilities in network protocols like ICMP (Internet Control Message Protocol), which is essential for services such as diagnostics and connectivity checks. Distinguishing between legitimate and malicious ICMP packets is challenging. TCP-land and Slowloris attacks involve sending a small number of packets to the target, often just a few per second, blending with legitimate traffic and making it harder to detect anomalies.
The \textcolor{black}{confusion matrix} of the tested methods on the Cloud dataset is shown in Table \ref{tab:confusio_matrix_cloud}, highlighting the results on the Ping-of-dead, Slowloris, and Tcp-land attacks. TAE outperforms MAE on these attacks. For instance, TAE correctly identifies 138 out of 145 samples for the Slowloris attack, while MAE identifies only 117. Similar results are observed for the other two attacks. Comparing TAE and Xgb, they are mostly equal except for the Slowloris attack, where TAE is slightly better. 

The final set of experiments examines the performance of TAE, MAE, and Xgb on the Mal3 dataset. The training set for Mal3 consists of normal traffic and a single attack type (DDoS), while the testing set includes various unknown attacks, such as C\&C-Heart-Beat-Attack, C\&C-Heart-Beat-File-Download, C\&C-Part-Of-A-Horizontal-Port-Scan, Okiru, and Part-Of-A-Horizontal-Port-Scan-Attack, as detailed in Table \ref{tab:confusio_matrix_mal3}.
TAE outperforms both MAE and Xgb in attack detection, correctly identifying 45,635 out of 47,694 attacks. In comparison, MAE and Xgb identified only 43,681 and 43,708, respectively. This highlights TAE's superior ability to detect unknown attacks in malware datasets.


\subsection{{\color{black}TAE's {\color {black} CATR} Compared to Codewords for RL}}

{\color{black}
We compare the results of representation learning models using TAE's {\color {black} CATR} and codewords, i.e., MLP and MAE. The codewords tested include one-hot encoding, Hadamard \cite{evron2023role} \cite{yang2015deep}, and the codeword of MAE (MAE's codeword), and the codewords are the means of data samples from different classes of TAE (TAE's means), as described in Algorithm \ref{algo:separated_multi_cate}.
Note that codewords are computed once before training, whereas {\color {black} CATRs} are updated for every data sample at each epoch during the training of the TAE model. Decision Trees (DT) are used to test all representation models. As observed in Table \ref{tab:result_codewords_inblack}, the accuracy and F1-score obtained by the representation learning model using TAE's {\color{black} CATR} outperform those using codewords, i.e., one-hot encoding, Hadamard, MAE's codeword, and TAE's means. In addition, the results for codewords using Hadamard \cite{evron2023role} \cite{yang2015deep} are better than those using one-hot encoding, while the results for codewords using TAE's means are better than those using one-hot encoding, Hadamard, and MAE's codewords. For example, the average accuracy obtained by TAE's {\color{black} CATR} over seven IoT datasets is 0.961, whereas the accuracies for MAE's codewords, one-hot encoding, Hadamard, and TAE means when applying MAE representation are 0.923, 0.897, 0.935, and 0.942, respectively. Interestingly, the representation models using TAE's means as fixed codewords rank second in terms of accuracy and F1-score. This is because TAE's means are generated from the input data and help separate data samples from different classes.
This helps explain the superior performance of TAE compared to other representation models that use fixed codewords.
}

\begin{table}[t] 
\caption{{\color{black} Comparison of the accuracy of the DT classifier using the original data and the data representations of MAE, CTVAE, and TAE on artificial datasets as the number of classes in the training set increases.}}
\label{tab:result-artificial-datasets-inblack}
\setlength\tabcolsep{1.5pt}
\centering
\scriptsize
\begin{tabular}{|c|c|c|c|c|c|c|c|c|c|c|c|}
\hline
{\color[HTML]{000000} No. Classes} & {\color[HTML]{000000} 50} & {\color[HTML]{000000} 100}& {\color[HTML]{000000} 150}& {\color[HTML]{000000} 200}& {\color[HTML]{000000} 250}& {\color[HTML]{000000} 300}& {\color[HTML]{000000} 350}& {\color[HTML]{000000} 400}& {\color[HTML]{000000} 450}& {\color[HTML]{000000} 500}& {\color[HTML]{000000} Avg}\\ \hline
{\color[HTML]{000000} DT} & {\color[HTML]{000000} 0.896} & {\color[HTML]{000000} 0.846} & {\color[HTML]{000000} 0.811} & {\color[HTML]{000000} 0.787} & {\color[HTML]{000000} 0.770} & {\color[HTML]{000000} 0.756} & {\color[HTML]{000000} 0.735} & {\color[HTML]{000000} 0.726} & {\color[HTML]{000000} 0.706} & {\color[HTML]{000000} 0.698} & {\color[HTML]{000000} 0.773} \\ \hline
{\color[HTML]{000000} MAE}& {\color[HTML]{000000} 0.725} & {\color[HTML]{000000} 0.636} & {\color[HTML]{000000} 0.592} & {\color[HTML]{000000} 0.503} & {\color[HTML]{000000} 0.407} & {\color[HTML]{000000} 0.258} & {\color[HTML]{000000} 0.165} & {\color[HTML]{000000} 0.140} & {\color[HTML]{000000} 0.142} & {\color[HTML]{000000} 0.119} & {\color[HTML]{000000} 0.369} \\ \hline
{\color[HTML]{000000} CTVAE} & {\color[HTML]{000000} 0.987} & {\color[HTML]{000000} 0.988} & {\color[HTML]{000000} 0.963} & {\color[HTML]{000000} 0.938} & {\color[HTML]{000000} 0.909} & {\color[HTML]{000000} 0.888} & {\color[HTML]{000000} 0.886} & {\color[HTML]{000000} 0.839} & {\color[HTML]{000000} 0.825} & {\color[HTML]{000000} 0.806} & {\color[HTML]{000000} 0.903} \\ \hline
{\color[HTML]{000000} TAE}& {\color[HTML]{000000} \textbf{0.998}} & {\color[HTML]{000000} \textbf{0.991}} & {\color[HTML]{000000} \textbf{0.977}} & {\color[HTML]{000000} \textbf{0.948}} & {\color[HTML]{000000} \textbf{0.912}} & {\color[HTML]{000000} \textbf{0.948}} & {\color[HTML]{000000} \textbf{0.901}} & {\color[HTML]{000000} \textbf{0.874}} & {\color[HTML]{000000} \textbf{0.860}} & {\color[HTML]{000000} \textbf{0.826}} & {\color[HTML]{000000} \textbf{0.923}} \\ \hline
\end{tabular}

\end{table}

\begin{table}[t] 
\caption{{\color{black} Accuracy comparison of DT classifier using original data and data representation of MAE, CTVAE, and TAE on small-sized artificial datasets.}}
\label{tab:result-artificial-datasets-small-size-inblack}
\setlength\tabcolsep{2pt}
\centering
\scriptsize

\begin{tabular}{|c|c|c|c|c|c|c|c|c|c|}
\hline
{\color[HTML]{000000} No. Classes} & {\color[HTML]{000000} 5} & {\color[HTML]{000000} 10}& {\color[HTML]{000000} 15}& {\color[HTML]{000000} 20}& {\color[HTML]{000000} 25}& {\color[HTML]{000000} 30}& {\color[HTML]{000000} 35}& {\color[HTML]{000000} 40}& {\color[HTML]{000000} Avg}  \\ \hline
{\color[HTML]{000000} DT}  & {\color[HTML]{000000} 0.953}& {\color[HTML]{000000} 0.903}& {\color[HTML]{000000} 0.862}& {\color[HTML]{000000} 0.843}& {\color[HTML]{000000} 0.845}& {\color[HTML]{000000} 0.844}& {\color[HTML]{000000} 0.806}& {\color[HTML]{000000} 0.781}& {\color[HTML]{000000} 0.855}\\ \hline
{\color[HTML]{000000} MAE} & {\color[HTML]{000000} 0.967}& {\color[HTML]{000000} 0.803}& {\color[HTML]{000000} 0.642}& {\color[HTML]{000000} 0.530}& {\color[HTML]{000000} 0.524}& {\color[HTML]{000000} 0.477}& {\color[HTML]{000000} 0.567}& {\color[HTML]{000000} 0.548}& {\color[HTML]{000000} 0.632}\\ \hline
{\color[HTML]{000000} CTVAE} & {\color[HTML]{000000} \textbf{1.000}} & {\color[HTML]{000000} 0.947}& {\color[HTML]{000000} 0.980}& {\color[HTML]{000000} 0.953}& {\color[HTML]{000000} 0.975}& {\color[HTML]{000000} 0.960}& {\color[HTML]{000000} 0.948}& {\color[HTML]{000000} 0.956}& {\color[HTML]{000000} 0.965}\\ \hline
{\color[HTML]{000000} TAE} & {\color[HTML]{000000} \textbf{1.000}} & {\color[HTML]{000000} \textbf{1.000}} & {\color[HTML]{000000} \textbf{1.000}} & {\color[HTML]{000000} \textbf{0.987}} & {\color[HTML]{000000} \textbf{0.991}} & {\color[HTML]{000000} \textbf{0.984}} & {\color[HTML]{000000} \textbf{0.987}} & {\color[HTML]{000000} \textbf{0.980}} & {\color[HTML]{000000} \textbf{0.991}} \\ \hline
\end{tabular}

\end{table}

\subsection{Scalability with Increasing Class Numbers}
We evaluate the performance of TAE as the number of classes in the training set increases.
Table \ref{tab:result-artificial-datasets-inblack} presents the accuracy achieved by Decision Trees (DT) using the original datasets, compared to the data representations of MAE, CTVAE, and TAE. Overall, the accuracy of the TAE data representation (TAE-DT) is significantly higher than that of the original data (DT) and the data representations of MAE (MAE-DT) and CTVAE (CTVAE-DT) across ten artificial datasets. For instance, the average accuracy obtained by TAE-DT across these datasets is 0.923, which is notably higher than the 0.773, 0.369, and 0.903 achieved by DT, MAE-DT, and CTVAE-DT, respectively. MAE-DT achieves the lowest accuracy because MAE struggles to address the problem as the number of classes in the training set increases.

We evaluate the performance of TAE on small-sized artificial datasets. As observed in Table \ref{tab:result-artificial-datasets-small-size-inblack}, the accuracy of the DT classifier using the data representation of TAE (TAE-DT) is significantly greater than that of MAE-DT, CTVAE-DT, and the original data (DT). For example, the average accuracy of DT, MAE-DT, CTVAE-DT, and TAE-DT is 0.855, 0.632, 0.965, and 0.991, respectively. The accuracy of DT on the original data decreases as the number of classes in the training set increases. Similarly, MAE-DT reports low accuracy as the number of classes in the training set increases.

\begin{table*}[!htp]
	\caption{\normalsize {\color{black} Comparision of performance of TAE, CTVAE, CTVAE using the transformation operator of TAE called CTVAE-tae, and TAE using  the transformation operator of CTVAE called TAE-ctvae.}}
	\label{tab:result_TAE_CTVAE_TAEctvae-CTVAEtae_black}
	\setlength\tabcolsep{3pt}
	\centering
	\footnotesize	
 \begin{tabular}{|c|c|ccccccc|cc|c|ccc|c|}
\hline
{\color[HTML]{000000} } & {\color[HTML]{000000} }& \multicolumn{7}{c|}{{\color[HTML]{000000} \textbf{IoT IDS}}} & \multicolumn{2}{c|}{{\color[HTML]{000000} \textbf{\begin{tabular}[c]{@{}c@{}}Network\\  IDS\end{tabular}}}}& {\color[HTML]{000000} \textbf{\begin{tabular}[c]{@{}c@{}}Cloud\\  IDS\end{tabular}}} & \multicolumn{3}{c|}{{\color[HTML]{000000} \textbf{Malware}}} & {\color[HTML]{000000} }\\ \cline{3-15}
\multirow{-2}{*}{{\color[HTML]{000000} \textbf{Metric}}} & \multirow{-2}{*}{{\color[HTML]{000000} \textbf{Method}}} & \multicolumn{1}{c|}{{\color[HTML]{000000} \textbf{IoT1}}}  & \multicolumn{1}{c|}{{\color[HTML]{000000} \textbf{IoT2}}} & \multicolumn{1}{c|}{{\color[HTML]{000000} \textbf{IoT3}}}  & \multicolumn{1}{c|}{{\color[HTML]{000000} \textbf{IoT4}}} & \multicolumn{1}{c|}{{\color[HTML]{000000} \textbf{IoT5}}}  & \multicolumn{1}{c|}{{\color[HTML]{000000} \textbf{IoT6}}}  & {\color[HTML]{000000} \textbf{IoT7}}  & \multicolumn{1}{c|}{{\color[HTML]{000000} \textbf{NSL}}}& {\color[HTML]{000000} \textbf{UNSW}}  & {\color[HTML]{000000} \textbf{Cloud}}& \multicolumn{1}{c|}{{\color[HTML]{000000} \textbf{Mal-1}}} & \multicolumn{1}{c|}{{\color[HTML]{000000} \textbf{Mal-2}}} & {\color[HTML]{000000} \textbf{Mal-3}} & \multirow{-2}{*}{{\color[HTML]{000000} \textbf{AVG}}} \\ \hline \hline
{\color[HTML]{000000} } & {\color[HTML]{000000} CTVAE} & \multicolumn{1}{c|}{{\color[HTML]{000000} \textbf{0.933}}} & \multicolumn{1}{c|}{{\color[HTML]{000000} 0.930}} & \multicolumn{1}{c|}{{\color[HTML]{000000} 0.965}}  & \multicolumn{1}{c|}{{\color[HTML]{000000} 0.939}} & \multicolumn{1}{c|}{{\color[HTML]{000000} 0.948}}  & \multicolumn{1}{c|}{{\color[HTML]{000000} 0.931}}  & {\color[HTML]{000000} 0.934}  & \multicolumn{1}{c|}{{\color[HTML]{000000} 0.863}}  & {\color[HTML]{000000} 0.741}  & {\color[HTML]{000000} 0.980} & \multicolumn{1}{c|}{{\color[HTML]{000000} 0.513}}  & \multicolumn{1}{c|}{{\color[HTML]{000000} 0.877}}  & {\color[HTML]{000000} 0.921}  & {\color[HTML]{000000} 0.883}  \\ \cline{2-16} 
{\color[HTML]{000000} } & {\color[HTML]{000000} CTVAE-tae}  & \multicolumn{1}{c|}{{\color[HTML]{000000} 0.923}}  & \multicolumn{1}{c|}{{\color[HTML]{000000} 0.925}} & \multicolumn{1}{c|}{{\color[HTML]{000000} 0.966}}  & \multicolumn{1}{c|}{{\color[HTML]{000000} 0.934}} & \multicolumn{1}{c|}{{\color[HTML]{000000} 0.975}}  & \multicolumn{1}{c|}{{\color[HTML]{000000} 0.926}}  & {\color[HTML]{000000} 0.952}  & \multicolumn{1}{c|}{{\color[HTML]{000000} 0.867}}  & {\color[HTML]{000000} 0.737}  & {\color[HTML]{000000} 0.980} & \multicolumn{1}{c|}{{\color[HTML]{000000} 0.517}}  & \multicolumn{1}{c|}{{\color[HTML]{000000} 0.892}}  & {\color[HTML]{000000} 0.922}  & {\color[HTML]{000000} 0.886}  \\ \cline{2-16} 
{\color[HTML]{000000} } & {\color[HTML]{000000} TAE-ctvae}  & \multicolumn{1}{c|}{{\color[HTML]{000000} 0.919}}  & \multicolumn{1}{c|}{{\color[HTML]{000000} \textbf{0.938}}}& \multicolumn{1}{c|}{{\color[HTML]{000000} 0.973}}  & \multicolumn{1}{c|}{{\color[HTML]{000000} \textbf{0.962}}}& \multicolumn{1}{c|}{{\color[HTML]{000000} 0.941}}  & \multicolumn{1}{c|}{{\color[HTML]{000000} 0.915}}  & {\color[HTML]{000000} 0.956}  & \multicolumn{1}{c|}{{\color[HTML]{000000} 0.854}}  & {\color[HTML]{000000} 0.736}  & {\color[HTML]{000000} 0.983} & \multicolumn{1}{c|}{{\color[HTML]{000000} 0.516}}  & \multicolumn{1}{c|}{{\color[HTML]{000000} 0.894}}  & {\color[HTML]{000000} 0.913}  & {\color[HTML]{000000} 0.885}  \\ \cline{2-16} 
\multirow{-4}{*}{{\color[HTML]{000000} accuracy}}  & \cellcolor[HTML]{CCCCCC}{\color[HTML]{000000} TAE}& \multicolumn{1}{c|}{\cellcolor[HTML]{CCCCCC}{\color[HTML]{000000} 0.931}}  & \multicolumn{1}{c|}{\cellcolor[HTML]{CCCCCC}{\color[HTML]{000000} 0.933}} & \multicolumn{1}{c|}{\cellcolor[HTML]{CCCCCC}{\color[HTML]{000000} \textbf{0.985}}} & \multicolumn{1}{c|}{\cellcolor[HTML]{CCCCCC}{\color[HTML]{000000} 0.949}} & \multicolumn{1}{c|}{\cellcolor[HTML]{CCCCCC}{\color[HTML]{000000} \textbf{0.993}}} & \multicolumn{1}{c|}{\cellcolor[HTML]{CCCCCC}{\color[HTML]{000000} \textbf{0.959}}} & \cellcolor[HTML]{CCCCCC}{\color[HTML]{000000} \textbf{0.975}} & \multicolumn{1}{c|}{\cellcolor[HTML]{CCCCCC}{\color[HTML]{000000} \textbf{0.873}}} & \cellcolor[HTML]{CCCCCC}{\color[HTML]{000000} \textbf{0.747}} & \cellcolor[HTML]{CCCCCC}{\color[HTML]{000000} \textbf{0.987}}& \multicolumn{1}{c|}{\cellcolor[HTML]{CCCCCC}{\color[HTML]{000000} \textbf{0.517}}} & \multicolumn{1}{c|}{\cellcolor[HTML]{CCCCCC}{\color[HTML]{000000} \textbf{0.914}}} & \cellcolor[HTML]{CCCCCC}{\color[HTML]{000000} \textbf{0.955}} & \cellcolor[HTML]{CCCCCC}{\color[HTML]{000000} \textbf{0.901}} \\ \hline \hline
{\color[HTML]{000000} } & {\color[HTML]{000000} CTVAE} & \multicolumn{1}{c|}{{\color[HTML]{000000} 0.915}}  & \multicolumn{1}{c|}{{\color[HTML]{000000} 0.916}} & \multicolumn{1}{c|}{{\color[HTML]{000000} 0.960}}  & \multicolumn{1}{c|}{{\color[HTML]{000000} 0.931}} & \multicolumn{1}{c|}{{\color[HTML]{000000} 0.937}}  & \multicolumn{1}{c|}{{\color[HTML]{000000} 0.927}}  & {\color[HTML]{000000} 0.923}  & \multicolumn{1}{c|}{{\color[HTML]{000000} 0.826}}  & {\color[HTML]{000000} 0.673} & {\color[HTML]{000000} 0.979} & \multicolumn{1}{c|}{{\color[HTML]{000000} 0.433}}  & \multicolumn{1}{c|}{{\color[HTML]{000000} 0.873}}  & {\color[HTML]{000000} 0.929}  & {\color[HTML]{000000} 0.863}  \\ \cline{2-16} 
{\color[HTML]{000000} } & {\color[HTML]{000000} CTVAE-tae}  & \multicolumn{1}{c|}{{\color[HTML]{000000} 0.895}}  & \multicolumn{1}{c|}{{\color[HTML]{000000} 0.908}} & \multicolumn{1}{c|}{{\color[HTML]{000000} 0.964}}  & \multicolumn{1}{c|}{{\color[HTML]{000000} 0.917}} & \multicolumn{1}{c|}{{\color[HTML]{000000} 0.974}}  & \multicolumn{1}{c|}{{\color[HTML]{000000} 0.923}}  & {\color[HTML]{000000} 0.940}  & \multicolumn{1}{c|}{{\color[HTML]{000000} 0.830}}  & {\color[HTML]{000000} \textbf{0.706}}  & {\color[HTML]{000000} 0.980} & \multicolumn{1}{c|}{{\color[HTML]{000000} 0.439}}  & \multicolumn{1}{c|}{{\color[HTML]{000000} 0.889}}  & {\color[HTML]{000000} 0.930}  & {\color[HTML]{000000} 0.869}  \\ \cline{2-16} 
{\color[HTML]{000000} } & {\color[HTML]{000000} TAE-ctvae}  & \multicolumn{1}{c|}{{\color[HTML]{000000} 0.893}}  & \multicolumn{1}{c|}{{\color[HTML]{000000} \textbf{0.926}}}& \multicolumn{1}{c|}{{\color[HTML]{000000} 0.971}}  & \multicolumn{1}{c|}{{\color[HTML]{000000} \textbf{0.958}}}& \multicolumn{1}{c|}{{\color[HTML]{000000} 0.926}}  & \multicolumn{1}{c|}{{\color[HTML]{000000} 0.903}}  & {\color[HTML]{000000} 0.947}  & \multicolumn{1}{c|}{{\color[HTML]{000000} 0.810}}  & {\color[HTML]{000000} 0.700}  & {\color[HTML]{000000} 0.981} & \multicolumn{1}{c|}{{\color[HTML]{000000} 0.439}}  & \multicolumn{1}{c|}{{\color[HTML]{000000} 0.891}}  & {\color[HTML]{000000} 0.922}  & {\color[HTML]{000000} 0.867}  \\ \cline{2-16} 
\multirow{-4}{*}{{\color[HTML]{000000} F1-score}} & \cellcolor[HTML]{CCCCCC}{\color[HTML]{000000} TAE}& \multicolumn{1}{c|}{\cellcolor[HTML]{CCCCCC}{\color[HTML]{000000} \textbf{0.916}}} & \multicolumn{1}{c|}{\cellcolor[HTML]{CCCCCC}{\color[HTML]{000000} 0.922}} & \multicolumn{1}{c|}{\cellcolor[HTML]{CCCCCC}{\color[HTML]{000000} \textbf{0.985}}} & \multicolumn{1}{c|}{\cellcolor[HTML]{CCCCCC}{\color[HTML]{000000} 0.944}} & \multicolumn{1}{c|}{\cellcolor[HTML]{CCCCCC}{\color[HTML]{000000} \textbf{0.993}}} & \multicolumn{1}{c|}{\cellcolor[HTML]{CCCCCC}{\color[HTML]{000000} \textbf{0.949}}} & \cellcolor[HTML]{CCCCCC}{\color[HTML]{000000} \textbf{0.970}} & \multicolumn{1}{c|}{\cellcolor[HTML]{CCCCCC}{\color[HTML]{000000} \textbf{0.844}}} & \cellcolor[HTML]{CCCCCC}{\color[HTML]{000000} 0.695}  & \cellcolor[HTML]{CCCCCC}{\color[HTML]{000000} \textbf{0.987}}& \multicolumn{1}{c|}{\cellcolor[HTML]{CCCCCC}{\color[HTML]{000000} \textbf{0.439}}} & \multicolumn{1}{c|}{\cellcolor[HTML]{CCCCCC}{\color[HTML]{000000} \textbf{0.912}}} & \cellcolor[HTML]{CCCCCC}{\color[HTML]{000000} \textbf{0.958}} & \cellcolor[HTML]{CCCCCC}{\color[HTML]{000000} \textbf{0.886}} \\ \hline
\end{tabular}

\end{table*}

\begin{figure*}[t]
    \centering
    \begin{tabular}{ccc} 
 \begin{subfigure}{0.25\textwidth}\centering\includegraphics[width=\linewidth]{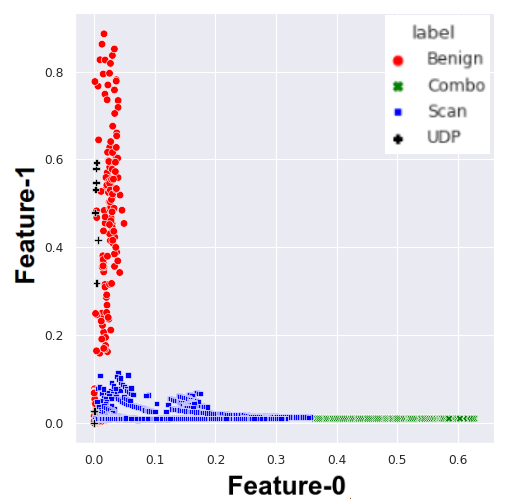}\end{subfigure} &
 \begin{subfigure}{0.25\textwidth}\centering\includegraphics[width=\linewidth]{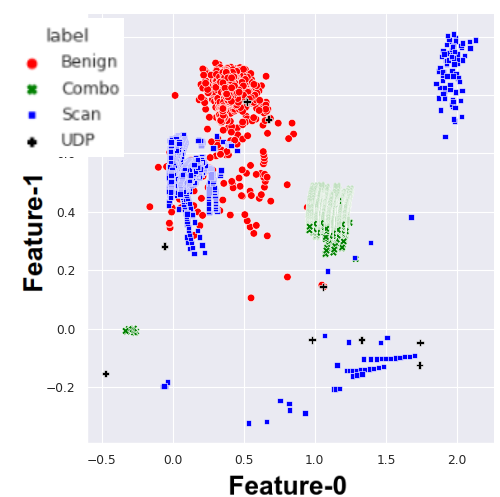}\end{subfigure} &
 \begin{subfigure}{0.25\textwidth}\centering\includegraphics[width=\linewidth]{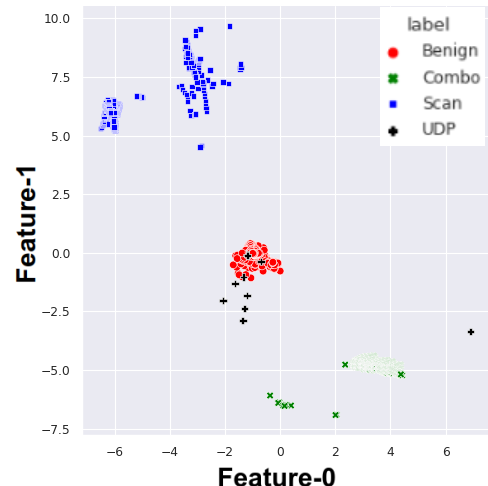}\end{subfigure} \\
 \textcolor{black} {a. Original data} & \textcolor{black} {b. Latent vector ($\mathbf{e}^{(i)}$)} & \textcolor{black} {c. Reconstruction representation ($\mathbf{z}^{(i)}$)}
    \end{tabular}
    \caption{Simulation of data representation using TAE on the IoT3 dataset.}
    \label{fig:dataRepresentionTAE}
\end{figure*}

{\color{black}
\subsection{Comparison with Knowledge Distillation}
}

\begin{table}[!htbp]
	\caption{\normalsize {\color{black} Performance comparison between the proposed TAE and KD with DeepMLP and ResNet teachers.}}
	\label{tab:KD_vs_TAE_acc}
	\setlength\tabcolsep{6pt}
	\centering
	\footnotesize	
	\begin{tabular}{|c|c|cccc|c|}
		\hline
		\multirow{2}{*}{\textbf{Scores}} & \multirow{2}{*}{\textbf{Method}} & \multicolumn{4}{c|}{\textbf{IoT IDS}}                                                                                            & \multirow{2}{*}{\textbf{Avg}} \\ \cline{3-6}
		&                                   & \multicolumn{1}{c|}{\textbf{IoT2}}  & \multicolumn{1}{c|}{\textbf{IoT3}}  & \multicolumn{1}{c|}{\textbf{IoT4}}  & \textbf{IoT5}  &                               \\ \hline
		\multirow{3}{*}{Acc}             & DeepMLP-KD  & \multicolumn{1}{c|}{0.906}          & \multicolumn{1}{c|}{0.955}          & \multicolumn{1}{c|}{\textbf{0.952}} & 0.956          & 0.942                         \\ \cline{2-7} 
		& RESNET-KD   & \multicolumn{1}{c|}{0.921}          & \multicolumn{1}{c|}{0.947}          & \multicolumn{1}{c|}{0.932}          & 0.941          & 0.935                         \\ \cline{2-7} 
		& TAE                               & \multicolumn{1}{c|}{\textbf{0.933}} & \multicolumn{1}{c|}{\textbf{0.985}} & \multicolumn{1}{c|}{0.949}          & \textbf{0.993} & \textbf{0.965}                \\ \hline
		\multirow{3}{*}{F-score}         & DeepMLP-KD  & \multicolumn{1}{c|}{0.872}          & \multicolumn{1}{c|}{0.969}          & \multicolumn{1}{c|}{\textbf{0.947}} & 0.949          & 0.934                         \\ \cline{2-7} 
		& RESNET-KD   & \multicolumn{1}{c|}{0.905}          & \multicolumn{1}{c|}{0.966}          & \multicolumn{1}{c|}{0.916}          & 0.926          & 0.928                         \\ \cline{2-7} 
		& TAE                               & \multicolumn{1}{c|}{\textbf{0.922}} & \multicolumn{1}{c|}{\textbf{0.985}} & \multicolumn{1}{c|}{0.944}          & \textbf{0.993} & \textbf{0.961}                \\ \hline
	\end{tabular}

\end{table}

\begin{table}[!htbp]
	\caption{\normalsize {\color{black} Comparison of conventional teacher--student KD and teacher-free TAE in terms of accuracy and model overhead on the IoT3 dataset.}}
	\label{tab:KD_vs_TAE}
	\setlength\tabcolsep{2.0pt}
	\centering
	\footnotesize	
	\begin{tabular}{|c|c|c|c|c|c|c|}
		\hline
    \textbf{Method} &
    \textbf{Teacher} &
    \textbf{\begin{tabular}[c]{@{}c@{}}Teacher\\ Size\end{tabular}} &
    \textbf{\begin{tabular}[c]{@{}c@{}}Client\\ Size\end{tabular}} &
    \textbf{\begin{tabular}[c]{@{}c@{}}\# Training\\ Params.\end{tabular}} &
    \textbf{\begin{tabular}[c]{@{}c@{}}\# Inference\\ Params.\end{tabular}} &
    \textbf{Accuracy} \\ \hline
		KD              & DeepMLP                & 418 KB                & 199 KB                                                               & 105.87K                                                                & 27.27K                                                                  & 0.955             \\ \hline
		KD              & ResNet                 & 852 KB                & 199 KB                                                               & 475.27K                                                                & 27.27K                                                                  & 0.969             \\ \hline
		TAE             & None                   & 0                     & 482 KB                                                               & \textbf{67.72K}                                                        & \textbf{22.54}K                                                                  & \textbf{0.985}    \\ \hline
	\end{tabular}
	
\end{table}
{\color{black}
	Table~\ref{tab:KD_vs_TAE_acc} shows that the proposed TAE consistently achieves higher average accuracy and F-score than both KD variants. Specifically, for TAE, the average accuracy and F-score are 0.965 and 0.961, respectively, compared with 0.942/0.934 for DeepMLP-based KD and 0.935/0.928 for ResNet-based KD. TAE also achieves the best performance on IoT2, IoT3, and IoT5, demonstrating its effectiveness over teacher-based KD approaches.
	
	Table~\ref{tab:KD_vs_TAE} compares TAE with conventional teacher--student KD using DeepMLP and ResNet teachers. TAE achieves the highest accuracy of 98.5\%, outperforming KD with DeepMLP and ResNet teachers, respectively, while eliminating the need for a teacher model. Moreover, TAE requires only 67.72K training parameters and 22.54K inference parameters, reducing the training parameter overhead by 36.0\% and 85.8\% compared with DeepMLP- and ResNet-based KD, respectively.
}

\section{TAE's CHARACTERISTICS}
\label{sec:TAE_analysis}
\subsection{{\color{black}Transformation Operator Comparision}}

{\color{black}
To further explain the result of TAE, we compare the results of TAE, CTVAE, and CTVAE using the transformation operator of TAE (referred to as CTVAE-tae), and TAE using the transformation operator of CTVAE (referred to as TAE-ctvae), as shown in Table \ref{tab:result_TAE_CTVAE_TAEctvae-CTVAEtae_black}.
Overall, the average accuracy and F1-score obtained by TAE are approximately 1.5\% higher than those of CTVAE, CTVAE-tae, and TAE-ctvae. This indicates that TAE effectively addresses the posterior collapse issue observed in CTVAE. Specifically, TAE achieves this by replacing the parameter trick and KL-divergence term in CTVAE's loss function with the generation of distinct {\color{black} CATR} based on the input data.
In addition, the average accuracy of CTVAE-tae is slightly higher than that of CTVAE, i.e., 0.886 vs. 0.883. This suggests that using the transformation operator of TAE is more effective than that of CTVAE.
}

\subsection{Representation Visualization}
\label{subsec:simulation}

This subsection analyzes representations of TAE by visualizing them in a 2D space. Fig. \ref{fig:dataRepresentionTAE} displays the distribution of the different vectors of TAE on the IoT3 dataset. To simplify the representation, the latent vector is set to two dimensions. Figs. \ref{fig:dataRepresentionTAE} (a), \ref{fig:dataRepresentionTAE} (b), and \ref{fig:dataRepresentionTAE} (c) show the simulation of the testing data. Fig. \ref{fig:dataRepresentionTAE} (a) shows the original data reduced to two dimensions by PCA. Fig. \ref{fig:dataRepresentionTAE} (b) shows the latent representation of TAE on the testing datasets, while Fig. \ref{fig:dataRepresentionTAE} (c) shows the reconstruction representation ($\textbf{y}^{(i)}$) on the testing datasets.

It can be observed that the original data samples in the testing datasets (see Fig. \ref{fig:dataRepresentionTAE} (a)) overlap. In the latent space, the data samples are still mixed, indicating that using the latent vector for downstream tasks may not be effective. However, most data samples in the reconstruction representation of TAE are well separated (Fig. \ref{fig:dataRepresentionTAE} (c)). For example, data samples from Benign and UDP in the original datasets overlap in the testing datasets, but they are much more separated in Fig. \ref{fig:dataRepresentionTAE} (c). Additionally, as seen in Fig. \ref{fig:dataRepresentionTAE} (c), most of the data samples in the reconstruction representation are shrunk to a small region, making it easier to classify the data samples of different classes.

\subsection {Representation Quality}
To further explain the superior performance of the TAE, we follow the idea from \cite{xanthopoulos2013linear} to evaluate the quality of the reconstruction representation in TAE. Specifically, we use two measures to quantify the representation quality, i.e., the average distance between means of classes, $d_{Bet}$, and the average distance between data samples with their mean $d_{Wit}$. Moreover, we also introduce a new harmonic measure called the \textit{representation-quality} that is calculated as $representation\mbox{-}quality = \frac{d_{Bet}}{d_{Wit}}$. A good representation is the one that  samples $\textbf{x}^{(i,c)}$ of class $c$ is closer to its mean $\mu^{(c)}$, whilst the distance between means of classes are as the largest. In other words, if the value of $representation\mbox{-}quality$ is greater, the better the representation is.

Table \ref{tab:data_quality} presents the representation quality of the original data (Orig) $\textbf{x}^{(i)}$ and the reconstruction representation $\hat{\textbf{z}}^{(i)}$ in TAE on the IoT datasets. It is clear that the value of $represenation\mbox{-}quality$ obtained by the reconstruction representation in TAE  is much greater than that obtained by the original data. This provides more pieces of evidence to explain the better performance obtained by the reconstruction representation of TAE compared to the original data.


\begin{table}[]
	\caption {Representation quality the original data the reconstruction representation.}
	\label{tab:data_quality}
	\setlength\tabcolsep{1.5pt}
	\fontsize{8pt}{9pt}\selectfont
	\centering
	\begin{tabular}{|c|cc|cc|cc|}
\hline
\multirow{2}{*}{\textbf{Data}} & \multicolumn{2}{c|}{\textbf{Between-class}}& \multicolumn{2}{c|}{\textbf{Within-class}}& \multicolumn{2}{c|}{\textbf{Representation quality}}\\ \cline{2-7} 
& \multicolumn{1}{c|}{\textbf{Orig}} & \textbf{TAE} & \multicolumn{1}{c|}{\textbf{Orig}}& \textbf{TAE}& \multicolumn{1}{c|}{\textbf{Orig}} & \textbf{TAE} \\ \hline
IoT1& \multicolumn{1}{c|}{1.1}  & \textbf{2.9} & \multicolumn{1}{c|}{\textbf{1.6E-3}} & 2.1E-3 & \multicolumn{1}{c|}{7.0E+02}& \textbf{1.4E+03} \\ \hline
IoT2& \multicolumn{1}{c|}{1.2}  & \textbf{4.1} & \multicolumn{1}{c|}{1.9E-3} & \textbf{704E-6} & \multicolumn{1}{c|}{6.5E+02}& \textbf{5.8E+03} \\ \hline
IoT3& \multicolumn{1}{c|}{1.1}  & \textbf{3.8} & \multicolumn{1}{c|}{2.1E-3} & \textbf{25E-6}  & \multicolumn{1}{c|}{5.5E+02}& \textbf{1.5E+05} \\ \hline
IoT4& \multicolumn{1}{c|}{1.2}  & \textbf{3.3} & \multicolumn{1}{c|}{2.9E-3} & \textbf{1.4E-3} & \multicolumn{1}{c|}{4.0E+02}& \textbf{2.4E+03} \\ \hline
IoT5& \multicolumn{1}{c|}{1.2}  & \textbf{3.7} & \multicolumn{1}{c|}{4.0E-3} & \textbf{61E-6}  & \multicolumn{1}{c|}{2.9E+02}& \textbf{6.0E+04} \\ \hline
IoT6& \multicolumn{1}{c|}{4.1}  & \textbf{5.2} & \multicolumn{1}{c|}{13.8E-3}& \textbf{1.6E-3} & \multicolumn{1}{c|}{3.0E+02}& \textbf{3.4E+03} \\ \hline
IoT7& \multicolumn{1}{c|}{3.2}  & \textbf{6.0} & \multicolumn{1}{c|}{9.8E-3} & \textbf{1.6E-3} & \multicolumn{1}{c|}{3.3E+02}& \textbf{3.8E+03} \\ \hline

\end{tabular}

\end{table}

\subsection{Reconstruction Representation vs Latent Pepresentation}

We conduct an experiment comparing the accuracy of the DT classifier using latent representation $\textbf{e}^{(i)}$ and reconstruction representation $\hat{\textbf{z}}^{(i)}$. The results in Table \ref{tab:TAE-encoder-vs-zhat} show that the accuracy from $\hat{\textbf{z}}^{(i)}$ is significantly higher than from $\textbf{e}^{(i)}$ across seven IoT datasets. This is due to the $\textbf{e}^{(i)}$ data samples of different classes being projected into small regions around means $\boldsymbol{\mu}^{(c)} $ (see Equation ($\ref{eq:loss_TAE}$)), resulting in frequent overlap, as illustrated in Fig.~\ref{fig:dataRepresentionTAE}. In contrast, $\hat{\textbf{z}}^{(i)}$ data samples of different classes are mapped to separate regions defined by $\boldsymbol{\hat{\mu}}^{(c)}$ (see Fig. \ref{fig:encoder-moved}). This demonstrates TAE's ability to transform the overlapping latent representation into a distinguishable representation, facilitating conventional machine learning in detecting cyberattacks.

\begin{figure}[t]
    \centering
    \begin{minipage}{0.22\textwidth}
 \centering
 \includegraphics[width=\linewidth]{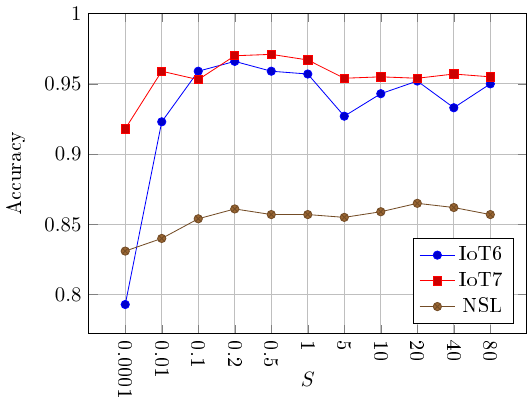}
 \caption*{a. Tuning parameter $S$}
    \end{minipage}
    \hspace{1em} 
    \begin{minipage}{0.22\textwidth}
 \centering
 \includegraphics[width=\linewidth]{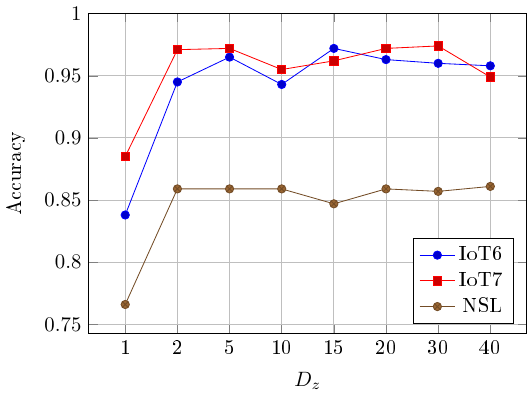}
 \caption*{b. Tuning parameter $D_z$}
    \end{minipage}
    \caption{Influence of accuracy of TAE based on $S$ and $D_z$.}
    \label{fig:tuning-params}
\end{figure}

\begin{table}[t]
	\caption{{\color{black} Performance of DT using the latent representation($\textbf{e}^{(i)}$) and the reconstruction representation (${\hat{\textbf{z}}}^{(i)}$)}}
	\label{tab:TAE-encoder-vs-zhat}
	\setlength\tabcolsep{2pt}
	\centering
	\scriptsize
    \begin{tabular}{|c|c|c|c|c|c|c|c|c|}
\hline
\textbf{Data} & \textbf{IoT1} & \textbf{IoT2}  & \textbf{IoT3}  & \textbf{IoT4}  & \textbf{IoT5}  & \textbf{IoT6}  & \textbf{IoT7}  & \textbf{Avg}   \\ \hline
$e^{(i)}$  & 0.883& 0.916& 0.941& 0.921& 0.953& 0.92 & 0.966& 0.929\\ \hline
$\hat{z}^{(i)}$      & \textbf{0.931}& \textbf{0.933} & \textbf{0.985} & \textbf{0.949} & \textbf{0.993} & \textbf{0.959} & \textbf{0.975} & \textbf{0.961} \\ \hline
\end{tabular}

\end{table}

\begin{table}[!t]
	\caption{{\color{black} Inference time and model size of MAE, CTVAE, and TAE on the IoT2 dataset.}}
	\label{tab:model_complexity-inblack}
	\setlength\tabcolsep{5pt}
	\centering
	\scriptsize
 \begin{tabular}{|c|c|c|c|}
\hline
{\color[HTML]{000000} \textbf{Metrics}}& {\color[HTML]{000000} \textbf{MAE}} & {\color[HTML]{000000} \textbf{CTVAE}} & {\color[HTML]{000000} \textbf{TAE}} \\ \hline
{\color[HTML]{000000} Model size (MB)}& {\color[HTML]{000000} 0.26}& {\color[HTML]{000000} 1.4}& {\color[HTML]{000000} 1.1} \\ \hline
{\color[HTML]{000000} Avg time (seconds)} & {\color[HTML]{000000} 2.6E-07}& {\color[HTML]{000000} 2.9E-07}  & {\color[HTML]{000000} 2.6E-07}\\ \hline
\end{tabular}

\end{table}



\subsection {TAE's Hyper-parameters} 
First, we investigate the influence of the hyper-parameter $S$ in Algorithm \ref{algo:separated_multi_cate} on the TAE's performance, as shown in Figure \ref{fig:tuning-params}a. The dimensionality of the representation vector $d_z$ is set at 10. When the value of $S$ is too small, the accuracy of the TAE obtained by the DT classifier decreases. This occurs because data samples of $\boldsymbol{\hat{\mu}}^{(c)}$ are closer together, as shown in Algorithm \ref{algo:separated_multi_cate} and Figure \ref{fig:encoder-moved}. For example, the accuracy is lower than 0.8 when $S$ is 0.0001, significantly below its average value of 0.95.
The influence of the latent space dimensionality $d_z$ is shown in Figure \ref{fig:tuning-params}b. The accuracy of TAE from the DT classifier declines when $d_z$ is too small. For instance, the accuracy on NSLKDD drops below 0.8 when $d_z = 1$, compared to an average of 0.86. With only 1 dimension, information is lost at the Encoder, reducing the DT classifier's accuracy. Additionally, the TAE model cannot operate when $d_z$ exceeds the input data's dimensionality due to the PCA in Algorithm \ref{algo:separated_multi_cate}.

\subsection{{\color{black}Model Complexity}}

{\color{black}
We present the experimental results on the complexity of the proposed model, TAE, in comparison to relevant works, i.e., MAE and CTVAE, on the IoT2 dataset. Note that the three models, i.e., MAE, CTVAE, and TAE, share the same neural network architecture for their encoder.
We measure the model size, i.e., the storage space occupied by the model on the hard disk (in MB), and the inference time, i.e., the average time required to process a single data sample, as reported in Table~\ref{tab:model_complexity-inblack}.
The table shows that all three models are relatively small, with sizes of approximately 1 MB. For MAE, CTVAE, and TAE, their inference time is nearly identical, around 0.26~$\mu$s, as they share the same complexity in the testing phase.
Thus, TAE is well-suited for cybersecurity applications and potentially for IoT systems, where a model size is approximately 1 MB.
}

\section{Conclusions}
\label{sec:conclusion}
{\color{black}

This paper proposed TAE, which learns discriminative class-wise representations without requiring an external teacher model. Theoretical analysis established conditions for perfect representation and characterized regimes in which TAE achieves lower empirical risk than methods using fixed class centers. Extensive experiments across 13 cybersecurity datasets demonstrated that TAE achieves up to 96.1\% average accuracy for IoT attack detection and 98.7\% for cloud intrusion detection. With a compact model size of approximately 1 MB and an average inference time of 0.26~$\mu$s per sample, TAE provides an effective and lightweight solution for cybersecurity and IoT applications.
}

In the future, our work can be extended in several directions. 
{\color{black}
The current {\color{black} CATR} can only be used with TAE's architecture/model and cannot be applied to other representation models, such as MLP and MAE. TAE's {\color{black} CATR} could be extended for use in other representation learning models. For example, one could develop a method to select key data points within a {\color{black} CATR} to form a group of codewords representing a class. To project a single data point into the group of codewords, one would need to project this data point alongside all the key data points within the group.
}
Second, the loss function of TAE requires a trade-off between its components. It would be beneficial to develop an approach that can automatically select optimal values for each dataset. Last but not least, the proposed model in this article has been evaluated on cybersecurity datasets. Therefore, it would be interesting to extend this model to other fields, such as computer vision and natural language processing.



\bibliographystyle{IEEEtran}
\bibliography{IEEEabrv,library}{}

\end{document}